\documentclass[a4paper,10pt]{article}
\pdfoutput=1

\usepackage{jheppub}
\usepackage{amsmath}
\usepackage{amssymb}
\usepackage{graphicx}
\usepackage[utf8]{inputenc}

\usepackage{color}
\definecolor{darkgreen}{rgb}{0,0.5,0}
\definecolor{darkblue}{rgb}{0,0,0.7}
\definecolor{darkred}{rgb}{0.5,0,0.0}
\definecolor{darkorange}{rgb}{0.8,0.4,0.0}

\newcommand{\ie}{i.e.\ }
\newcommand{\eg}{e.g.\ }
\newcommand{\zcut}{\ensuremath{z_{\text{cut}}}}

\makeatletter
\g@addto@macro\bfseries{\boldmath}
\makeatother

\makeatletter
\newcommand{\oset}[3][0ex]{%
  \mathrel{\mathop{#3}\limits^{
    \vbox to#1{\kern-2\ex@
    \hbox{$\scriptstyle#2$}\vss}}}}
\makeatother

\title{\sf Computing $N$-subjettiness for boosted jets}
\author[a]{Davide Napoletano}
\author[a]{and Gregory Soyez}
\affiliation[a]{IPhT, CEA Saclay, CNRS, Universit\'e Paris-Saclay, F-91191 Gif-sur-Yvette cedex, France}

\emailAdd{davide.napoletano@ipht.fr}
\emailAdd{gregory.soyez@ipht.fr}


\abstract{
  Jet substructure tools have proven useful in a number of high-energy
  particle-physics studies. A particular case is the discrimination,
  or tagging, between a boosted jet originated from an electroweak
  boson (signal), and a standard QCD parton (background).
  A common way to achieve this is to cut on a measure of the radiation
  inside the jet, \ie a jet shape.
  Over the last few years, analytic calculations of jet substructure
  have allowed for a deeper understanding of these tools and for the
  development of more efficient ones.
  However, analytic calculations are often limited to the region where
  the jet shape is small.
  In this paper we introduce a new approach in perturbative QCD to
  compute jet shapes for a generic boosted jets, waiving the above
  limitation.
  We focus on an example common in the substructure literature: the
  jet mass distribution after a cut on the $N$-subjettiness
  $\tau_{21}$ ratio, extending previous works to the
  region relevant for phenomenology.
  We compare our analytic predictions to Monte Carlo simulations for
  both plain and SoftDrop-groomed jets.
  We use our results to construct analytically a decorrelated tagger.
}

\begin{document}

\maketitle

\section{Introduction}

The field of jet substructure, \ie the use and study of the internal
dynamic properties of jets, has gained a sizeable importance at the
LHC over the past few years, both theoretically and experimentally
(see e.g.~\cite{Larkoski:2017jix} and~\cite{Asquith:2018igt} for
recent reviews).
The main application of jet substructure is likely the tagging of
highly boosted electroweak ($H/W/Z$) bosons or top quarks, produced
with transverse momenta much larger than their mass, a situation which
appears increasingly often at the LHC, in particular in searches for
new physics
(e.g.~\cite{Sirunyan:2018omb,Sirunyan:2018hsl,CMS:2018tuj,Aaboud:2018juj,Aaboud:2018zba,Aaboud:2018mjh})
and studies of the Higgs boson~\cite{Sirunyan:2017dgc}.
It has also seen many more recent developments, noticeably analytic
studies of substructure observables (e.g.~\cite{Dasgupta:2013ihk,Dasgupta:2013via,Larkoski:2013eya,Larkoski:2014wba,Larkoski:2015kga,Dasgupta:2015lxh,Dasgupta:2016ktv,Larkoski:2017cqq,Moult:2017okx,Dasgupta:2018emf}),
the use of substructure techniques to probe the quark-gluon plasma in
high-energy heavy-ion collisions
(e.g.~\cite{Sirunyan:2017bsd,Caffarri:2017bmh,Chien:2018dfn,AndrewsALICEQM18,Connors:2017ptx}),
the use of Machine Learning techniques
(e.g.~\cite{Cogan:2014oua,deOliveira:2015xxd,Komiske:2016rsd,Kasieczka:2017nvn,Louppe:2017ipp,Egan:2017ojy,Andreassen:2018apy,Datta:2017rhs,Datta:2017lxt,Komiske:2017aww,Dreyer:2018nbf}),
and standard model measurements
(e.g.~\cite{CMS:2017tdn,Aaboud:2017qwh}) alongside precision
calculations in QCD (e.g.~\cite{Marzani:2017mva,Marzani:2017kqd,
  Frye:2016aiz}).

When tagging boosted bosons, many jet substructure observables are
based on two basic observations: (i) electroweak bosons tend to have
multiple high-energy cores/subjets --- two for an electroweak boson, 3
for a top quark ---, and (ii) the QCD radiation patterns is different
in a boosted boson compared to a standard QCD jet.
Modern substructure taggers used by the LHC experiments combine
both ideas.
Several tools exploiting the first idea have already been studied
analytically~\cite{Dasgupta:2013ihk,Dasgupta:2013via,Larkoski:2014wba,Dasgupta:2016ktv,Dasgupta:2018emf},
sometimes even targeting
precision~\cite{Marzani:2017mva,Marzani:2017kqd, Frye:2016aiz}.
Observables relying on radiation patterns, typically by imposing a cut
on a jet shape, are more complicated to study. As we discuss in this work,
this is primarily because they require at least 3 particles in the
jet, meaning that they start one order later in the perturbative QCD
series expansion relative to tools of the first category and usually
involve additional scales.
Currently, several jet shapes have been studied in the limit where
both the jet mass --- more precisely the ratio of the jet mass over
its transverse momentum, $m/p_t$ --- and the cut on the shape are
small~\cite{Dasgupta:2015lxh}.
Recent calculations have been performed in Soft-Collinear Effective
Theory (SCET) for some energy correlation functions,
$D_2$~\cite{Larkoski:2015kga,Larkoski:2017cqq,Moult:2017okx}, still
imposing $m\ll p_t$ but not requiring specific conditions on $D_2$.

The idea behind this paper is to push the level of our analytic
understanding of another jet shape, the $N$-subjettiness
ratio~\cite{Thaler:2010tr,Kim:2010uj,Thaler:2011gf},
$\tau_{21}=\tau_2/\tau_1$, in a purely perturbative QCD approach.
To do that, we extend the calculation of~\cite{Dasgupta:2015lxh},
done in the limit where both $m\ll p_t$ and $\tau_{21}\ll 1$, to the
situation where we only require $m\ll p_t$ and allow $\tau_{21}$ to
take any value.
We choose to focus on the $\tau_{21}$ ratio (with the $\beta$
parameter set to 2) mainly because the structure of the calculation
greatly simplifies. However, we believe that the same method can be
applied to a series of other jet shapes, like $\tau_{21}$ with any
value of $\beta$, energy correlation functions, dichroic ratios or
even shapes relevant for top tagging, which are left for a future
work.
Additionally, the method presented below essentially amounts to
computing a three-jet observable in the two-jet limit, a notoriously
complicated situation to address in the context of resummation. We
therefore hope that our results have an impact beyond the field of jet
substructure.

Besides the obvious interest in understanding the internal properties
of jets from a first-principles viewpoint, expanding our analytic
knowledge of jet substructure observables has two potential benefits:
it can lead to the introduction of better tools (see
e.g.\cite{Dasgupta:2013ihk,Larkoski:2014gra,Dasgupta:2016ktv,Salam:2016yht})
and it can lead to precision measurements in the context of
standard-model studies.
We give an example of the former by constructing a decorrelated
tagger~\cite{Dolen:2016kst} based on our analytic results.

This paper is organised as follows in Sec.~\ref{sec:defs} we
briefly review the definitions of $N$-subjettiness we use throughout
this paper. In Sec.~\ref{sec:target-accuracy} we summarise the findings
from the earlier study which are of relevance for this paper and
discuss the improved accuracy which we target in this paper.
Our main findings are presented in Sec.~\ref{sec:calc-finite-tau}
first performing the calculation for the double-differential
distribution in both the jet mass and the $\tau_{21}$ ratio, then
addressing the case, more relevant for phenomenological applications,
of the mass distribution with a cut on $\tau_{21}$.
Comparisons to numerical Monte Carlo simulations are presented in
Sec.~\ref{sec:mc-tests} before we conclude in
Sec.~\ref{sec:conclusions}.

\section{$N$-subjettiness}\label{sec:defs}

For a given jet and a set of $N$ axes $a_1,\dots,a_N$,
$N$-subjettiness is defined as
\begin{equation}\label{eq:tauN}
  \tau_N^{(\beta)} = \frac{1}{p_t R^\beta} \sum_{i\in{\rm jet}}
  p_{ti} \min(\theta_{ia_1}^\beta,\dots,\theta_{ia_N}^\beta),
\end{equation}
where the sum runs over all the constituents of the jet, of momentum
$p_{ti}$ and with an angular distance $\theta_{ia_j}$ to the axis
$a_j$.
$\beta$ is a parameter of $N$-subjettiness and in what follows we will
concentrate on the case $\beta=2$. The main reason for this choice is
that it simplifies the calculation in the case of the jet mass. Also
$\beta=2$ shows better performance in Monte Carlo studies than the
more standard choice $\beta=1$, with the main drawback that the former
is more sensitive to non-perturbative effects than the latter. This
can be addressed by lightly grooming the jet, e.g. with
SoftDrop~\cite{Larkoski:2014wba} before computing $N$-subjettiness
(see Sec.~3.2 of~\cite{Bendavid:2018nar} for a systematic study).

There are several ways to specify the axes. Common choices include
using exclusive $k_t$ axes or using ``minimal'' axes, \ie the axes
that minimise $\tau_N$. Here, we will either consider the minimal axes
or the case of exclusive axes obtained after re-clustering the jet
with the generalised-$k_t$ algorithm with $p=1/2$.\footnote{For a
  generic $\beta$, one could use the generalised $k_t$ algorithm with
  $p=1/\beta$.}
The motivation behind this choice has been explained
in~\cite{Dasgupta:2015lxh}: the ordering in $p_{ti} \theta_i^2$
corresponding to $\beta=2$ $N$-subjettiness in Eq.~(\ref{eq:tauN}) is
preserved by the clustering, at least in the strongly-ordered
limit.
To the accuracy we target in this paper, the generalised-$k_t$ and
minimal axes are equivalent.\footnote{Note however that the default
  implementation of the minimal axes ({\tt{MultiPass\_Axes}}) in the
  {\tt{fjcontrib}}~\cite{fjcontrib} $N$-subjettiness code starts with
  the $k_t$ axes as a seed. In cases where only a small number of particles
  are present, the
  $k_t$ and generalised-$k_t$($1/2$) axes differ significantly --- \eg
  in cases with 2 soft emissions with $z_1\theta_1^2\gg z_2\theta_2^2$
  and $z_1\theta_1\ll 2_2\theta_2$ --- and the code sometimes fails to
  find the right minimum.
  An easy workaround is to use instead the
  {\tt{MultiPass\_Manual\_Axes}}, setting manually the seed axes to the
  generalised-$k_t$($1/2$) axes. This is what we use in this
  paper.}
We discuss in more detail the extent to which the two 
choices of axes are equivalent in Sec.~\ref{sec:mc-tests-fo}.

\section{Targeted accuracy and hints from previous
  studies}\label{sec:target-accuracy}

Our calculation aims at including two regimes: the leading
($\tau_{21}$-dependent) logarithms of the jet mass relevant in the
boosted limit, $\rho=m^2/(p_tR)^2\ll 1$, and the leading (double)
logarithms of $\tau_{21}$ in the limit where $\tau_{21}\ll 1$.
In the boosted limit, the dominant contribution to jet mass
distributions comes from double logarithms of $\rho$,
corresponding to contributions of the form $\alpha_s^n\log^n(\rho)$.
These terms arise from the constraints on the jet mass distribution
and are independent of $\tau_{21}$.
For $\tau_{21}\ll 1$, the extra $N$-subjettiness constrain brings new
double-logarithmic terms of the form
$\alpha_s\log(1/\rho)\log(1/\tau)$ and $\alpha_s\log^2(1/\tau)$ which
have to be resummed to all orders.
This was done in~\cite{Dasgupta:2015lxh} and we briefly review these
results later in this Section.
In this paper, we are instead interested in the region where the
$N$-subjettiness constraint is not necessarily small.
In the logarithmic expansion in $\rho$, once the
$\tau_{21}$-independent double-logarithms of $\rho$ have been
extracted, the leading terms affected by the $N$-subjettiness
constraint are single-logarithmic terms in $\rho$, in the form
of $\alpha_s^n\log^n(1/\rho)f_n(\tau_{21})$ with $f_n$ to be
determined.
The main novelty of this paper is to compute these contributions, \ie
the exact form of the $f_n(\tau_{21})$ coefficients, while keeping the
full double-logarithmic structure (in both $\rho$ and $\tau_{21}$) in the
small $\tau_{21}$ limit.
This last point means that beyond the (single-logarithmic in $\rho$)
contribution to $f_n(\tau_{21})$ proportional to $\log^n(\tau_{21})$,
we also want to resum terms enhanced by double-logarithms of
$\tau_{21}$, $\alpha_s^n\log^{2n}(\tau_{21})$. In other words, our
accuracy includes both the leading (single-logarithmic) terms in
$\rho$ at any $\tau_{21}$ as well as the double-logarithmic in either
$\rho$ or $\tau_{21}$ relevant in the small $\tau_{21}$ limit.

Before turning to the full computation, let us first review the
computation from~\cite{Dasgupta:2015lxh}, in the limit
$\rho\ll 1$, $\tau_{21}\ll 1$.
We focus on the jet mass distribution with a
cut on $\tau_{21}<\tau$:\footnote{Unless explicitly stated
  otherwise, $\tau_{21}$ will denote a specific $N$-subjettiness value
  and $\tau$ will refer to a cut on $\tau_{21}$.}
\begin{equation}\label{eq:cumulative-distribution}
  H(\rho, <\tau) \equiv
  \left.\frac{\rho}{\sigma}\frac{d\sigma}{d\rho}\right|_{\tau_{21}<\tau}.
\end{equation}
In the double-logarithmic approximation (in both $\log(1/\rho)$ and
$\log(1/\tau)$), emissions in the jet can be considered
strongly-ordered in $z_i\theta_i^2$, with $z_i$ the transverse
momentum fraction of emission $i$ and $\theta_i$ its emission
angle.\footnote{From now on, we use a notation for which
  angles are normalised to $R$, \ie the actual emission angle is
  $\theta R$.}  We can therefore assume
\begin{equation}\label{eq:strong-mass-ordering}
  z_1\theta_1^2\gg z_2\theta_2^2\gg \dots \gg z_n\theta_n^2,
\end{equation}
as well as a strong angular ordering between the emissions.
In that case, the jet mass is dominated by the first emission. For
$\tau_N$, the $N$ axes will align with the ``leading'' parton and the
$N-1$ first emissions so that $\tau_N$ is dominated by the
$N^{\text{th}}$ emission. In our case, we therefore have
\begin{equation}\label{eq:tau21-ordered}
  \rho\approx \tau_1\approx z_1\theta_1^2
  \qquad\text{ and }\quad
  \tau_2\approx z_2\theta_2^2.
\end{equation}
With this at hand, the leading-logarithmic mass distribution can be
written as
\begin{equation}\label{eq:dist-ll-plain}
  H_{LL}(\rho, <\tau) = \int_0^1\frac{d\theta_1^2}{\theta_1^2}\,dz_1\,P(z_1)
    \frac{\alpha_s(z_1\theta_1 p_tR)}{2\pi}\,
    \rho\delta(z_1\theta_1^2-\rho)\,
  e^{-R_{\text{plain}}(\rho) - R_\tau(\tau;\rho,z_1)},
\end{equation}
with (the first expression below is introduced for later convenience)
\begin{align}
  R'_{\text{plain}}(\rho) & = \int_0^1\frac{d\theta^2}{\theta^2}\,dz\,P(z)
    \frac{\alpha_s(z\theta p_tR)}{2\pi}\,
    \rho\delta(z\theta^2-\rho)
    \overset{\text{f.c.}}{=} \frac{\alpha_sC_R}{\pi}\big[\log(1/\rho)+B_i\big],\label{eq:Rp_plain}\\
  R_{\text{plain}}(\rho)
  & = \int_0^1\frac{d\theta^2}{\theta^2}\,dz\,P(z)
    \frac{\alpha_s(z\theta p_tR)}{2\pi}\,
    \Theta(z\theta^2>\rho)
    \overset{\text{f.c.}}{=}
    \frac{\alpha_sC_R}{2\pi}\big[\log(1/\rho)+B_i\big]^2,\label{eq:R_plain}\\
  R_\tau(\tau;\rho,z_1) 
  & = R_{\text{plain}}(\tau\rho) - R_{\text{plain}}(\rho) +
    \int_0^{\theta_1^2}\frac{d\theta_{12}^2}{\theta_{12}^2}\int_0^1\,dz\,P(z)
    \frac{\alpha_s(zz_1\theta_{12} p_tR)}{2\pi}\,
    \Theta(z(\theta_{12}/\theta_1)^2>\tau)\nonumber\\
  & \overset{\text{f.c.}}{=}
    \frac{\alpha_sC_R}{2\pi}\left[2(\log(1/\rho)+B_i)\log(1/\tau)+\log^2(1/\tau)\right]
    + \frac{\alpha_sC_A}{2\pi}\big[\log(1/\tau)+B_g\big]^2.\label{eq:Rtau_plain}
\end{align}
Eq.~(\ref{eq:dist-ll-plain}) shows that $H(\rho,<\tau)$ receives 3
contributions: (i) a contribution from the real emission ``1'' which
dominates the jet mass, if it were not for the explicit dependence of
$R_\tau$ on $z_1$, this integration would lead to an overall
$R'_{\text{plain}}$ factor given by Eq.~(\ref{eq:Rp_plain}); (ii) a
Sudakov factor $\exp[-R_{\text{plain}}(\rho)]$, given by
Eq.~(\ref{eq:R_plain}), associated with the jet mass vetoing real
emissions with $z\theta^2>z_1\theta_1^2$; and (iii) a Sudakov factor
$\exp(-R_\tau)$, Eq.~(\ref{eq:Rtau_plain}), associated with the cut on
$\tau_{21}$, imposing that there are no additional real emissions with
$z_1\theta_1^2>z\theta^2>z_2\theta_2^2$.
The results indicated by ``f.c.'' in the expressions above have been
obtained assuming a fixed-coupling approximation to
highlight the logarithms that arise in the various contributions to
$H(\rho,<\tau)$.
For completeness, results for the radiators used throughout this paper
are given in Appendix~\ref{app:radiators}.
Finally, unless explicitly mentioned otherwise, we use a {\it
  modified} leading-logarithmic approximation to compute the
radiators, i.e. include the dominant leading logarithms as well as the
correction coming from hard collinear splittings, as explicit in the
fixed-coupling expressions above. In practice, we obtain this by
replacing the splitting function $P(z)$ by
$\tfrac{2C_R}{z}\Theta(\log(z)<B_i)$, with $C_R$ the appropriate
colour factor and $B_i=B_q$ (resp. $B_g$) introducing the contribution from
hard-collinear splittings for quarks (resp. gluons).
Note that $R_\tau(\tau;\rho)$ also includes a contribution,
proportional to $C_A$, corresponding to {\it secondary} emissions,
\ie to the situation where the emission which dominates $\tau_2$ is
emitted from emission ``1'' which dominates $\tau_1$.
In the end, the physical interpretation of the above result is that,
on top of the plain jet mass distribution
$R'_{\text{plain}}(\rho)\, \exp[-R_{\text{plain}}(\rho)]$, we gain an
extra exponential suppression, $\exp[-R_\tau(\tau;\rho)]$, due to the
constraint on $N$-subjettiness.
Note that since $R_\tau(\tau;\rho)$ depends on $z_1$ due to the
running of $\alpha_s$, the integration over $z_1$ in
(\ref{eq:dist-ll-plain}) --- which would otherwise give a
$R'_{\text{plain}}(\rho)$ factor --- has to be kept explicit.

While Eq.~(\ref{eq:dist-ll-plain}) captures the main physics
ingredients observed in Monte Carlo simulations, it is not without
limitations.
First, one can show that the signal events would also have a Sudakov
suppression factor. This means that one does not want to take the
$\tau_{21}$ cut too small. This motivates the calculation of the finite
$\tau_{21}$ corrections to (\ref{eq:dist-ll-plain}), for which
we introduce a generic powerful method next Section.
Second, Monte Carlo studies show --- see also
Sec.~\ref{sec:mc-tests} --- that the $\tau_{21}$ and mass
distributions are significantly affected by initial-state radiation
and non-perturbative effects.
One can obtain much more robust distributions by grooming the jet
prior to imposing the constraint on $\tau_{21}$, albeit
at a small cost in performance.
We will therefore also consider the case of jets groomed with the
modified MassDrop Tagger~\cite{Dasgupta:2013ihk} or
SoftDrop~\cite{Larkoski:2014wba}.
The above calculation remains valid, up to a redefinition of its
basic pieces:
\begin{equation}\label{eq:dist-ll-SD}
  H_{\text{LL,SD}}(\rho, <\tau) = \int_0^1\frac{d\theta_1^2}{\theta_1^2}\,dz_1\,P(z_1)
    \frac{\alpha_s(z_1\theta_1 p_tR)}{2\pi}\,\Theta(z_1>\zcut\theta_1^\beta)\,
    \rho\delta(z_1\theta_1^2-\rho)\,
  e^{-R_{\text{SD}}(\rho) - R_{\tau,\text{SD}}(\tau;\rho,z_1)},
\end{equation}
with
\begin{align}
  R'_{\text{SD}}(\rho) & = \int_0^1\frac{d\theta^2}{\theta^2}\,dz\,P(z)
    \frac{\alpha_s(z\theta p_tR)}{2\pi}\,\Theta(z>\zcut\theta^\beta)\,
    \rho\delta(z\theta^2-\rho),\label{eq:Rp_SD}\\
  R_{\text{SD}}(\rho)
  & = \int_0^1\frac{d\theta^2}{\theta^2}\,dz\,P(z)
    \frac{\alpha_s(z\theta p_tR)}{2\pi}\,\Theta(z>\zcut\theta^\beta)\,
    \Theta(z\theta^2>\rho),\label{eq:R_SD}\\
  R_{\text{SD}}(\tau;\rho,z_1) 
  & = \int_0^1\frac{d\theta^2}{\theta^2}\,dz\,P(z)
    \frac{\alpha_s(z\theta
    p_tR)}{2\pi}\,\Theta(z>\zcut\theta^\beta\text{ or }\theta<\theta_1)\,
    \Theta(\rho>z\theta^2>\rho\tau)\nonumber\\
  &+ \int_0^{\theta_1^2}\frac{d\theta_{12}^2}{\theta_{12}^2}\int_0^1\,dz\,P(z)
    \frac{\alpha_s(zz_1\theta_{12} p_tR)}{2\pi}\,
    \Theta(z(\theta_{12}/\theta_1)^2>\tau)\label{eq:Rtau_SD}
\end{align}
where the results are presented for SoftDrop with a generic
$\zcut$ and $\beta$ and one can obtain expressions for the
mMDT by setting $\beta$ to 0.
Compared to the plain-jet case this implies 
a cut on $z$ such that $z>\zcut\theta^\beta$. 
The only exception is the extra contribution $\theta<\theta_1$ present
in the definition of $R_{\tau,\text{SD}}$. This comes from the fact
that if emission ``1'' is the first to trigger the mMDT/SoftDrop
condition, then the mMDT/SD declustering procedure stops and all
emission at angles smaller than $\theta_1$ are kept in the groomed
jet.

\section{Calculation for finite $\tau$ cut}\label{sec:calc-finite-tau}

We now turn to the main calculation of this paper: the inclusion of
the finite-$\tau$ contributions to $H(\rho;<\tau)$. 
Compared to the previous section where, in the strongly-ordered limit,
$\tau_{21}=(z_2\theta_2^2)/(z_1\theta_1^2)$, a finite
$\tau_{21}$ implies $z_1\theta_1^2\gtrsim z_2\theta_2^2$.
More generally, this means that to perform a calculation at finite
$\tau_{21}$ we need to lift the ordering assumption between the
emissions in a jet, \ie we have\footnote{Emissions with much smaller
  values of $z\theta^2$ do not significantly contribute to either
  $\tau_1$ or $\tau_2$ and are therefore irrelevant.} 
\begin{equation}\label{eq:no-mass-ordering}
  \rho_1\sim \rho_2\sim \dots \sim \rho_n,
  \qquad\text{with }\rho_i = z_i\theta_i^2.
\end{equation}
With no specific ordering in mass (\ie in $\rho_i$), the dominant
logarithmic behaviour will come from a series of emissions strongly
ordered in angle, so we can assume in what follows that
\begin{equation}\label{eq:strong-angular-ordering}
  \theta_1\ll \theta_2\ll \dots \ll \theta_n,
\end{equation}
or, equivalently, a strong ordering in momentum fraction
\begin{equation}\label{eq:strong-energy-ordering}
  z_1\gg z_2\gg \dots \gg z_n.
\end{equation}
This ordering yields a coefficient of the form
$\alpha_s^n\log^n(\rho)f_n(\tau)$ where the $n$ powers of $\log(\rho)$
come from the strong ordering in angle and the $\tau$-dependent
coefficient $f_n(\tau)$ has to be computed.

The situation with no mass ordering and the strong angular ordering is
reminiscent of what one considers when computing multiple-emission
corrections to the jet mass, contributing at NLL, single-logarithmic,
accuracy. The main difference here is the addition of a constraint on
$N$-subjettiness.
This analogy suggests that one can use CAESAR-like
techniques~\cite{Banfi:2004yd} to compute the distribution
$H(\rho,<\tau)$. In what follows, we show how to do this in three
steps: first, we find a generic expression for $\tau_{21}$ based on a
set of $n$ emissions satisfying the
constraints~(\ref{eq:no-mass-ordering})
and~(\ref{eq:strong-angular-ordering}); then in
Sec.~\ref{sec:diff-xsect} we show in details how to derive an
expression for the double-differential cross-section $d^2\sigma/d\rho
d\tau_{21}$ before considering the case of $H(\rho,<\tau)$ in
Sec.~\ref{sec:cumul-distrib}. The reason to begin with the
double-differential distribution is that it is technically a bit
simpler than the cumulative distribution $H$ allowing us to focus on
the generic ideas behind the calculation.

\subsection{Computing $N$-subjettiness for a given set of
  emissions}\label{sec:tau-value}

The first thing we need is an expression for $\tau_{21}$ computed from
a set of emissions satisfying~(\ref{eq:no-mass-ordering})
and~(\ref{eq:strong-angular-ordering}).
In our small $\rho$ limit, the mass and $\tau_1$ coincide and are
known to be given by
\begin{equation}\label{eq:rho-tau-expression}
  \rho = \tau_1 = \sum_{i=1}^n \rho_i.
\end{equation}

As for the calculation of $\tau_2$, we start by investigating the 
case of minimal axes
For this, we consider a given partition of the emissions 
into two subjets. The emissions at large angle in a subjet are also
the softest, meaning that, up to negligible recoil
corrections,\footnote{For $\beta=1$ $N$-subjettiness, the recoil would
  not be negligible unless one works with a recoil-free axis like the
  one obtained with the winner-takes-all recombination
  scheme~\cite{Larkoski:2014uqa}.} the axes will be aligned with the
hardest particle in each of the 2 subjets. It is therefore sufficient
to consider the cases where one of the axes is aligned with the parent
hard particle and the second axes is aligned with one of the other
particles, say $j$. In that case, all particles with $i<j$ are
clustered with the parent particle, and all particles with $i>j$ are
clustered either with the parent particle or with particle $j$. That
means that, assuming a second axis aligned with emission $j$, we get
\begin{equation}
  \tau_2^{(j)}
  \simeq \sum_{i=1}^{j-1} z_i \theta_i^2
       + \sum_{i=j+1}^{n} z_i \min(\theta_{ij}^2,\theta_i^2) 
  \simeq \sum_{i=1}^{j-1} \rho_i
       + \sum_{i=j+1}^{n} \rho_i
  \simeq \tau_1 - \rho_j,
\end{equation}
where, for the second equality, we assume $\theta_{i} \simeq\theta_{ij}$
for $i>j$ which follows from strong angular ordering.
By definition of the minimal axes, we still need to choose the $j$
that minimises $\tau_2^{(j)}$, and this is simply the $j$ that gives the
largest $\rho_j$, yielding to
\begin{equation}\label{eq:tau21}
  \tau_2 \simeq \tau_1 - \max_i \rho_i,\qquad\text{and}\quad
  \tau_{21} \simeq 1-\frac{\max_i \rho_i}{\rho}.
\end{equation}

The same kind of arguments can be applied with
generalised-$k_t(p=1/2)$ clustering. At each step of the clustering
the minimal distance is either $d_{i0}=z_i\theta_i^2$ (with the
index ``0'' referring to the leading parton) or
$d_{ij}=z_i\theta_{ij}^2$ with $i>j$. Since in that case
$\theta_{ij}\sim \theta_i$ and $z_j\gg z_i$ one will simply cluster
the particle $i$ with the smallest $\rho_i=z_i\theta_i^2$ either with
the parent particle or with a particle $j$ with $j<i$, without
affecting the kinematics of the particle one clusters with.
This is iterated until the last step where one clusters the particle
with the largest $\rho_i$.  The expression for $\tau_{21}$ in the
generalised-$k_t(p=1/2)$ case is therefore the same as in the minimal
one.

Eq.~(\ref{eq:tau21}) has an interesting structure: for a set of
$n$ emissions, the maximal value of $\tau_{21}$ on can reach is
$\tau_{21}=\tfrac{n-1}{n}$, achieved when $\rho_1=\dots=\rho_n=\tfrac{\rho}{n}$, \ie
when all the emissions contribute equally to the jet mass.
One should therefore expect transition points at
$\tau_{21}=\tfrac{1}{2},\tfrac{2}{3},\tfrac{3}{4},\dots$.
Also, since each additional emission comes with an extra factor of
$\alpha_s$ --- accompanied by a logarithm of the jet mass as we shall
see below --- crossing one of these thresholds requires going further
in the perturbative expansion, with a transition point at
$\tau_{21}=\tfrac{1}{2}$ at leading order, at $\tau_{21}=\tfrac{2}{3}$
at NLO, etc...

Note that although the condition~(\ref{eq:no-mass-ordering}) that we
use to derive our expression for $\tau_{21}$ differs from the strong
ordering in mass, Eq.~(\ref{eq:strong-mass-ordering}), use in the
small $\tau_{21}$ limit~\cite{Dasgupta:2015lxh}, the two expressions
coincide in the small $\tau_{21}$ limit. In other words, we can use
Eq.~(\ref{eq:tau21}) for small $\tau_{21}$.

\begin{figure}[!t]
  \includegraphics[width=0.48\textwidth,page=1]{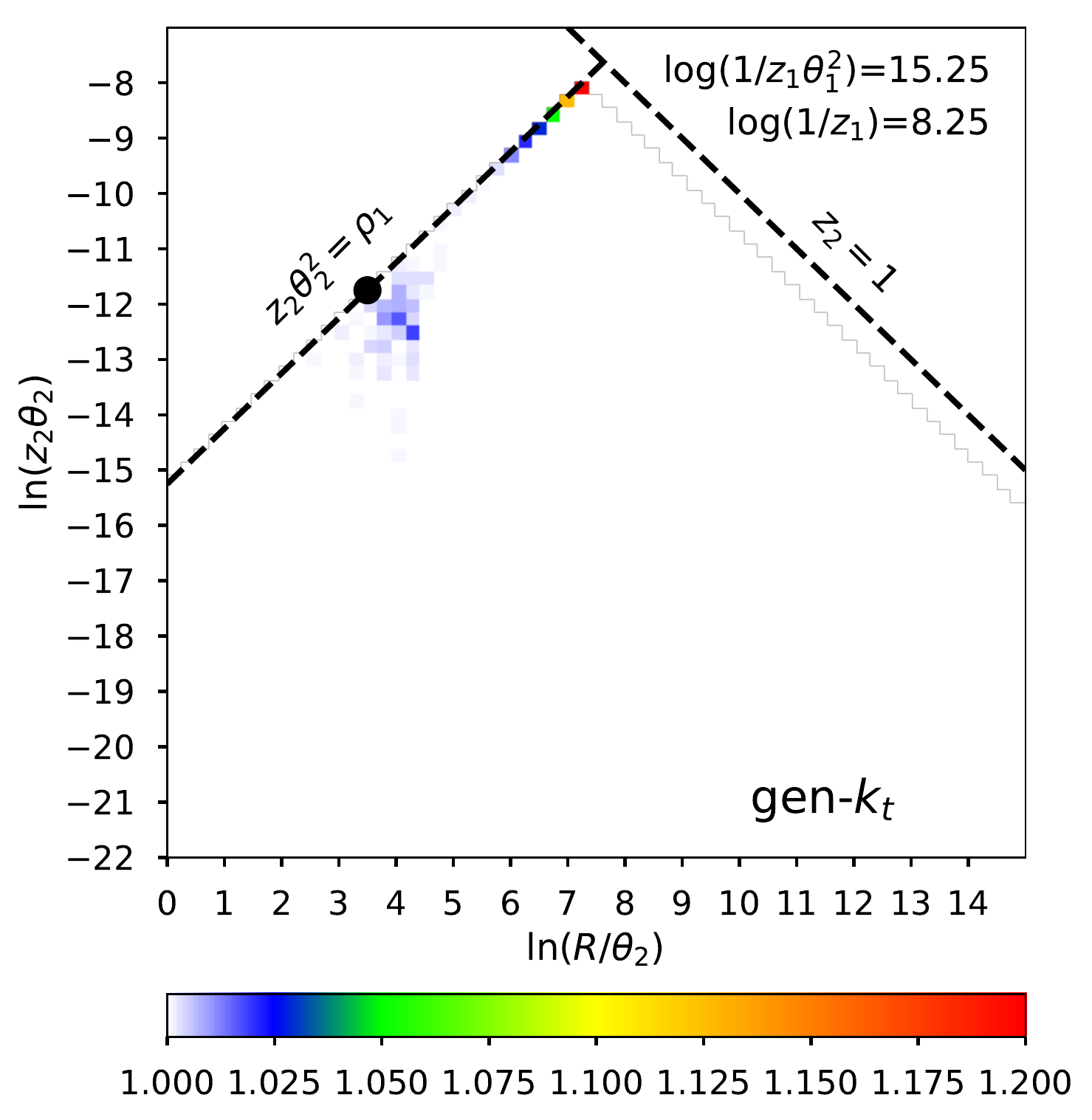}%
  \hfill%
  \includegraphics[width=0.48\textwidth,page=2]{figs/lund.pdf}%
  \caption{Plots of the ratio of the $\tau_{21}$ obtained in different
    limits. A first emission is fixed at the indicated point and the
    plots are done varying the second emission. On the left, we show
    the ratio
    $\tau_{21}^{\text{(gen-$k_t$)}}/\tau_{21}^{\text{(min)}}$ and on
    the right the ratio
    $\tau_{21}^{\text{(soft+ang.-ordered)}}/\tau_{21}^{\text{(min)}}$.}\label{fig:lund}
\end{figure}

To check the validity of (\ref{eq:tau21}) 
we consider events with 3 particles defined as follows:
\begin{align}
  p_{t0} &= (1-z_1-z_2) p_t, & y_0&=0        & \phi_0&=0, \nonumber\\
  p_{t1} &= z_1 p_t,  & y_1&=\theta_1, & \phi_1&=0,\nonumber \\
  p_{t2} &= z_2 p_t,  & y_2&=\theta_2\cos(\varphi), & \phi_2&=\theta_2\sin(\varphi).\label{eq:3-particle-config}
\end{align}
Fixing $\theta_1$ and $z_1$, varying $\theta_2$ and $z_2$, and
averaging over $\varphi$, we compute the value of $\tau_{21}$ in three
cases.  The first one, which we use as reference in the following, is
computed using the minimisation procedure as implemented in
{\tt{fjcontrib}}~\cite{fjcontrib}. A second case is considered when we
take the generalised-$k_t$ definition for the axes. The last one is
given by our approximation, which we dub ``soft+ang.-ordered'', implying
that it is obtained from the previous ones by taking their soft limit
and strong angular-ordering
(cf. Eq.~(\ref{eq:strong-angular-ordering})).
In order to check the level of agreement of these three different
definitions, we plot in Fig.~\ref{fig:lund} ratios of the value of
$\tau_{21}$ obtained in either the generalised-$k_t$ or the
``soft+ang.-ordered'' cases over the minimal axes value, as a function
of $\log(R/\theta_2)$ and $\log(z_2\theta_2)$, keeping
$z_2<\tfrac{1}{2}$ and $\rho_2=z_2\theta_2^2<\rho_1=z_1\theta_1^2$.

First, we see that the generalised-$k_t$ axes are in very good
agreement with the minimal axes. For $\rho_2\lesssim\rho_1$, we see
small deviations for large $z_2$ or $\theta_2\approx\theta_1$ and we
discuss this further in Sec.~\ref{sec:mc-tests-fo}.
Then, we see that our approximation, Eq.~(\ref{eq:tau21})
overestimates the minimal $\tau_{21}$ in two regions: at large $z_2$
and for $\theta_2\approx\theta_1$. Again, we discuss the influence of
these regions in Sec.~\ref{sec:mc-tests-fo} but the key point here is
that they are both of finite width, therefore not giving leading
logarithmic contributions.

\subsection{Differential $\tau_{21}$ distribution}\label{sec:diff-xsect}
We start the presentation of our results with the double differential
distribution in $\rho$ and $\tau_{21}$:\footnote{Throughout this
  paper, we compute distributions for a fixed jet mass $\rho$. The
  $\tau_{21}$ distribution with no constraints on the jet mass is
  infrared unsafe. Nevertheless, it remains
  ``Sudakov-safe''~\cite{Larkoski:2013paa,Larkoski:2015lea}.}
\begin{equation}\label{eq:differential-distribution}
  f(\rho,\tau_{21}) \equiv \frac{\rho\tau_{21}}{\sigma}\frac{d^2\sigma}{d\rho d\tau_{21}}.
\end{equation}
We do this, although our final goal is to compute the distribution $H(\rho,<\tau)$
with a cut on $\tau_{21}$,
as this hides some of the technical details in that case, while 
presenting all the main steps needed for the method presented in this
work.
We also leave aside for the moment secondary emissions, which contribute
at the double-logarithmic accuracy in $\tau_{21}$ but are not
enhanced by logarithms of $\rho$.
At the targeted accuracy, it is sufficient to consider any number $n$
of independent real gluon emissions, strongly ordered in angle (or in
momentum fraction, cf. Eqs.~(\ref{eq:strong-angular-ordering}) and
(\ref{eq:strong-energy-ordering})), dressed with virtual corrections.
This can be written as 
\begin{equation}\label{eq:differential-start}
  f(\rho,\tau)
  = \lim_{\epsilon\to 0}e^{-\int_\epsilon^1 d\omega_v}\sum_{n=1}^\infty\frac{1}{n!}
  \int_\epsilon^1 \prod_{i=1}^n d\omega_i\,
  \rho\delta\Big(\rho-\sum_{i=1}^n \rho_i\Big)\,
  \tau\delta\Big(\tau-1+\frac{\max_i\rho_i}{\rho}\Big),
\end{equation}
with the shorthand notation
\begin{equation}\label{eq:phase-space-shorthand}
\int_\epsilon d\omega_i \equiv \int_{\rho_i>\epsilon}
\frac{d\theta_i^2}{\theta_i^2}\,dz_i\,P(z_i)\frac{\alpha_s(z_i\theta_i p_tR)}{2\pi}
\end{equation}
The exponential pre-factor corresponds to virtual corrections, as made
explicit by the subscript $v$, and the two $\delta$ correspond to the
constraints on the jet mass and $N$-subjettiness ratio. 
We use Eq.~(\ref{eq:tau21}) to compute the value of $\tau_{21}$, and
we replace $\tau_{21}$ with $\tau$, to keep the notation more compact
throughout this section, unless otherwise explicitly stated.
Since the constraints only involve $\rho_i$, we can simplify our
phase-space integration and write
\begin{equation}\label{eq:phase-space-shorthand-simple}
\int_\epsilon d\omega_i = \int_\epsilon \frac{d\rho_i}{\rho_i} R'(\rho_i),
\end{equation}
with $R'(\rho_i)\equiv R'_\text{plain}(\rho_i)$ given by
Eq.~(\ref{eq:Rp_plain}), showing explicitly that each emission is
enhanced by a logarithm of the jet mass.
The next step is to single out the emission with the largest
$\rho_i$. Calling this emission $\rho_a$ and relabelling the remaining
emissions $\rho_1,\dots,\rho_p$, with $p=n-1$, one gets
\[
  f(\rho,\tau)
  = \int_0^1\frac{d\rho_a}{\rho_a}R'(\rho_a)\lim_{\epsilon\to 0}e^{-\int_\epsilon^1 \frac{d\rho_v}{\rho_v}R'(\rho_v)}\sum_{p=1}^\infty\frac{1}{p!}
  \int_\epsilon^{\rho_a} \prod_{i=1}^p \frac{d\rho_i}{\rho_i}R'(\rho_i)\,
  \rho\delta\Big(\rho-\rho_a-\sum_{i=1}^p \rho_i\Big)\,
  \tau\delta\Big(\tau-1+\frac{\rho_a}{\rho}\Big).
\]
In the above equation, we explicitly impose that each of the
$\rho_i$ has to be smaller than $\rho_a$.
The constraint on $\tau$ can be used to perform the $\rho_a$
integration,
\begin{equation}\label{eq:differential-after-rhoa}
  f(\rho,\tau)
  = \frac{\tau}{1-\tau}R'((1-\tau)\rho)\lim_{\epsilon\to 0}e^{-\int_\epsilon^1 \frac{d\rho_v}{\rho_v}R'(\rho_v)}\sum_{p=1}^\infty\frac{1}{p!}
  \int_\epsilon^{(1-\tau)\rho} \prod_{i=1}^p \frac{d\rho_i}{\rho_i}R'(\rho_i)\,
  \rho\delta\Big(\rho\tau-\sum_{i=1}^p \rho_i\Big).
\end{equation}

At this stage, we have to distinguish two cases: $\tau<1-\tau$ (\ie
$\tau<\tfrac{1}{2}$) and $\tau>1-\tau$ (\ie $\tau>\tfrac{1}{2}$).
When $\tau<\tfrac{1}{2}$, the constraint $\sum_i\rho_i=\rho\tau$
implies that each of the individual $\rho_i$ is smaller than
$\rho\tau$ hence the upper integration boundary $(1-\tau)\rho$ is
irrelevant.
Physically, this means that the appropriate scale for all of the
$\rho_i$ is $\rho\tau$.
We then define rescaled variables $\xi_i=\rho_i/(\tau\rho)$ and
$\varepsilon=\epsilon/(\tau\rho)$.  Within our accuracy, we replace
$R'(\rho_i)$ by $R'(\tau\rho)$ and rewrite the virtual corrections as
\begin{equation}
  e^{-\int_\epsilon^1 \frac{d\rho_v}{\rho_v}R'(\rho_v)}
   = e^{-R(\rho\tau)-R'(\rho\tau)\log(1/\varepsilon)},
\end{equation}
where we used $R$ from Eq.~(\ref{eq:R_plain}).
Eq.~(\ref{eq:differential-after-rhoa}) thus becomes
\begin{align}\label{eq:differential-small-tau}
  f(\rho,\tau)
  & \overset{\tau<1/2}=  \frac{R'((1-\tau)\rho)}{1-\tau}e^{-R(\rho\tau)}
    \lim_{\varepsilon\to 0} \sum_{p=1}^\infty\frac{R'^p(\rho\tau)}{p!}
  \int_\varepsilon^1 \prod_{i=1}^p \frac{d\xi_i}{\xi_i}\,
    e^{-R'(\rho\tau)\log(1/\varepsilon)}\delta\Big(1-\sum_{i=1}^p\xi_i\Big),
    \nonumber\\
  & \overset{\tau<1/2}= \frac{R'((1-\tau)\rho)R'(\rho\tau)}{1-\tau}
    \frac{e^{-R(\rho\tau)-\gamma_ER'(\rho\tau)}}{\Gamma(1+R'(\rho\tau))},
\end{align}
with $\gamma_E$ the Euler-Mascheroni constant.

This result is remarkably simple: the factor
$\tfrac{e^{-\gamma_ER'(\rho\tau)}}{\Gamma(1+R'(\rho\tau))}$ is the
standard expectation for the single-logarithmic multiple-emission
contribution to additive observables, in the limit of small
$\rho\tau$.
In particular, we note that Eq.~(\ref{eq:differential-small-tau})
includes the resummation of the terms enhanced by a double-logarithm
of $\tau$, modulo the contribution from secondary emissions that we
discuss at the end of this section.
We stress that the key point is to realise that the appropriate scale
for the $\rho_i$ emissions in~(\ref{eq:differential-after-rhoa}) is
$\rho\tau$.\footnote{On a technical side, we note the scale
  $\varepsilon$ (after rescaling) should be taken to satisfy
  $\rho\tau\ll\varepsilon\ll 1$, \ie such that
  $\log(1/\varepsilon)\ll\log(1/\rho\tau)$, cf. \eg
  \cite{Banfi:2004yd}, which is allowed since our observable is
  recursively infrared-and-collinear safe.}
Note that all finite $\tau$ effects are captured by
the pre-factor $\frac{1}{1-\tau}$.

The case of $\tau>\tfrac{1}{2}$ is a bit more delicate since one now
has to enforce the constraint $\rho_i<(1-\tau)\rho$. In this case
$(1-\tau)\rho$ becomes the appropriate physical scale for the $\rho_i$
and we now define the rescaled variables
$\zeta_i=\rho_i/((1-\tau)\rho)$. Using the same method as above leads
to
\begin{align}
  f(\rho,\tau)
  & \overset{\tau>1/2}=  R'((1-\tau)\rho)\frac{\tau}{(1-\tau)^2}e^{-R((1-\tau)\rho)}\nonumber\\
  &\phantom{\overset{\tau>1/2}=}\quad
    \lim_{\varepsilon\to 0} \sum_{p=1}^\infty\frac{R'^p((1-\tau)\rho)}{p!}
  \int_\varepsilon^1 \prod_{i=1}^p \frac{d\xi_i}{\xi_i}\,
    e^{-R'((1-\tau)\rho)\log(1/\varepsilon)}\delta\Big(\frac{\tau}{1-\tau}-\sum_{i=1}^p\zeta_i\Big).
\end{align}
We have not been able to perform analytically the integration over the
set of rescaled emissions $\zeta_i$ for a generic value of
$\tfrac{\tau}{1-\tau}$. We solve this problem by defining a {\it
  multiple-emission function}
\begin{equation}\label{eq:fME}
  \frac{e^{-\gamma_ER'}}{\Gamma(R')}f_{\text{ME}}(x;R')
  = \lim_{\varepsilon\to 0}
  \sum_{n=1}^\infty\frac{R'^n}{n!}
  \prod_{i=1}^n\int_{\varepsilon}^1
  \frac{dx_i}{x_i}\,
  e^{-R'\log(1/\varepsilon)} \delta(x-\sum_{i=1}^n x_i).
\end{equation}
so that the $N$-subjettiness distribution can then be
written as
\begin{equation}\label{eq:differential-large-tau}
  f(\rho,\tau)
  \overset{\tau>1/2}= R'^2((1-\tau)\rho))\frac{\tau}{(1-\tau)^2}
  \frac{e^{-R((1-\tau)\rho)-\gamma_ER'((1-\tau)\rho)}}{\Gamma(1+R'((1-\tau)\rho))}f_\text{ME}\Big(\frac{\tau}{1-\tau};R'((1-\tau)\rho)\Big),
\end{equation}

\begin{figure}[!t]
  \includegraphics[width=0.48\textwidth,page=1]{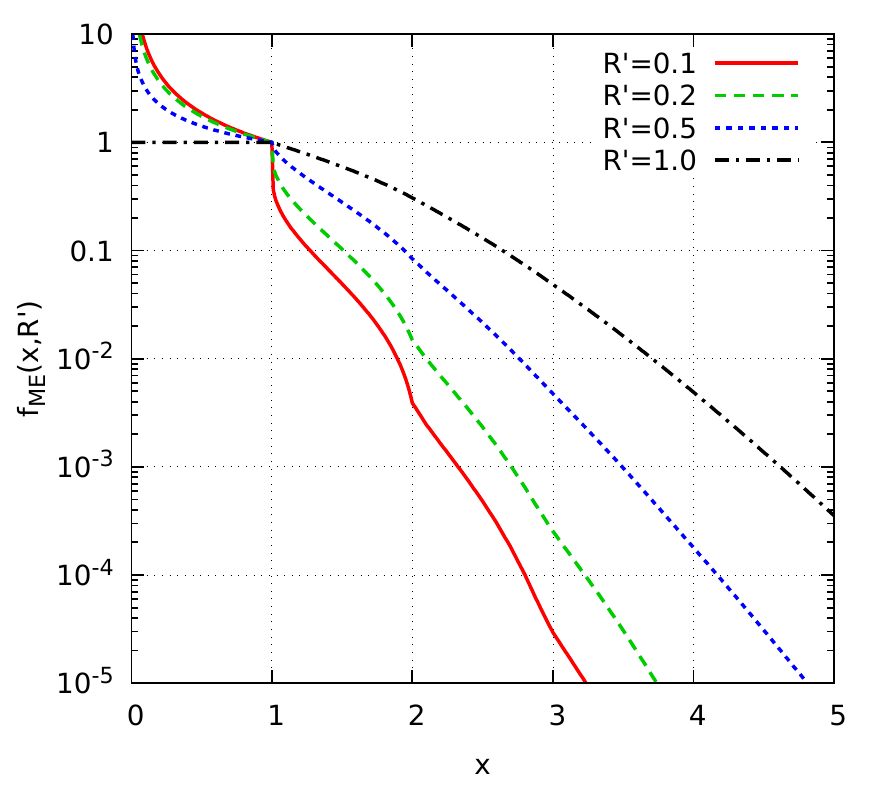}%
  \hfill%
  \includegraphics[width=0.48\textwidth,page=2]{figs/fme.pdf}
  \caption{Plot of $f_{\text{ME}}(x,R')$ as a function of $x$ for
    several representative values of $R'$ in both logarithmic (left)
    and linear (right) scales.}\label{fig:fme}
\end{figure}
Some analytic results for $f_{\text{ME}}$ are given in
Appendix~\ref{app:fme}, although in general it can be computed
numerically for any value of $x$ and $R'$.
To picture the main features of Eq.~(\ref{eq:differential-large-tau}), 
we plot $f_{\text{ME}}$ in Fig.~\ref{fig:fme}. 
Firstly, we can see that $f_{\text{ME}}(1,R')=1$,
which means that $f(\rho,\tau)$,
Eqs.~(\ref{eq:differential-small-tau})
and~(\ref{eq:differential-large-tau}), is continuous at
$\tau=\tfrac{1}{2}$.
One then sees a relatively fast decreases of $f_{\text{ME}}$ with $x$.
Furthermore, the plot shows that, especially for small $R'$,
$f_{\text{ME}}(x,R')$ has kinks at integer values of $x$. These
directly correspond to the transition points at $\tau=\tfrac{n-1}{n}$
mentioned at the end of Sec.~\ref{sec:tau-value}, as well as to the
fact that $f_{\text{ME}}(x\ge n,R')$ requires at least $n+1$ emissions
in the jet.
The transition point at $x=1$ is particularly visible, and it
implies that we expect a shoulder in the $\tau_{21}$ distribution at
$\tau=\frac{1}{2}$.

Finally, we need to take into account the fact that, at small $\tau$,
one would get an additional double logarithmic contribution in $\tau$
coming from secondary emissions, i.e. emissions from ``$\rho_a$''
which are enhanced by double logarithms of $\tau$ when
$\rho_{i=1,\dots,p}\ll\rho_a$.
This would add an extra Sudakov suppression to the
above result if it were not for running-coupling corrections to
secondary emissions (which explicitly depend on the $k_t$ scale
$z_a\theta_ap_tR$ of the emission ``$\rho_a$'') that dominates the mass
(cf.~(\ref{eq:Rtau_plain})). The integration over $z_a$ therefore has
to be kept explicit by writing
\begin{equation}
  R'((1-\tau)\rho) = \int_{(1-\tau)\rho}^1 d z_a\,P(z_a)\,\frac{\alpha_s(\sqrt{z_a(1-\tau)\rho}p_tR)}{2\pi}.
\end{equation}
One then has to add a $C_A$ contribution to $R(\rho\tau)$ above which
becomes (with $\theta_a^2=(1-\tau)\rho/z_a$)
\begin{align}
  R(\rho\tau;z_a) & = \int_0^1\frac{d\theta^2}{\theta^2}\,dz\,P(z)
    \frac{\alpha_s(z\theta p_tR)}{2\pi}\,
    \Theta(z\theta^2>\rho\tau)\\
   & + \int_0^{\theta_a^2}\frac{d\theta_{12}^2}{\theta_{12}^2}\int_0^1\,dz\,P(z)
    \frac{\alpha_s(zz_a\theta_{12} p_tR)}{2\pi}\,
    \Theta\Big(z\Big(\frac{\theta_{12}}{\theta_a}\Big)^2>\frac{\tau}{1-\tau}\Big)\label{eq:Rrhotau}
\end{align}
and similar for $R'(\rho\tau;z_a)$.

Note that the integral of the distribution $f(\rho,\tau)$ over
$\tau$, only equals the expected resummed differential mass
distribution
\[
\frac{\rho}{\sigma} \frac{d\sigma}{d\rho} =
R'(\rho)\frac{e^{-R(\rho)-\gamma_E R'(\rho)}}{\Gamma(1+R'(\rho))}
\]
up to subleading $\alpha_s$ corrections. This is expected given
the approximations made in the calculation, and has been explicitly
checked numerically.

Finally, note that the factors of $1-\tau$ appearing in
the $R$ and $R'$ factors go beyond our accuracy and we could simply
replace $R'((1-\tau)\rho)$ by $R'(\rho)$ and $R((1-\tau)\rho)$ by
$R(\rho)-\log(1-\tau)R'(\rho)$. This however introduces a
discontinuity at $\tau=\tfrac{1}{2}$ which, albeit beyond our
accuracy, may not be desired. Other options, all valid within our
accuracy (while still maintaining continuity) include replacing the scale
$(1-\tau)\rho$ by either $\rho/2$ or $\tau\rho$ for
$\tau>\tfrac{1}{2}$, or replacing the scale $\tau\rho$ by $2\tau\rho$
for $\tau<\tfrac{1}{2}$, or using $\tfrac{\tau}{1-\tau}\rho$ for
$\tau<\tfrac{1}{2}$ and simply $\rho$ for $\tau>\tfrac{1}{2}$ (always
including the appropriate single-logarithmic expansion for $R$).

\subsection{Cumulative $\tau$ distribution}\label{sec:cumul-distrib}

The calculation of the cumulative distribution
$H(\rho,\tau_{21}<\tau)$, which we use in all the subsequent
comparisons to Monte Carlo simulations, follows
closely that presented in the previous section for the double
differential case, up to a few extra minor technicalities.
The first difference is that we now impose a cut on $\tau$ instead
of taking it at a fixed value, \ie we replace
\[
  \tau\delta\Big(\tau-1+\frac{\max_i\rho_i}{\rho}\Big)
  \qquad\text{ by }\qquad
  \Theta\Big(1-\frac{\max_i\rho_i}{\rho}<\tau\Big).
\]
As before, we single out the emission with the largest $\rho_i$
to obtain
\begin{align}
  H(\rho,<\tau)
  = \int_0^1\frac{d\rho_a}{\rho_a}R'(\rho_a)\lim_{\epsilon\to 0} & e^{-\int_\epsilon^1 \frac{d\rho_v}{\rho_v}R'(\rho_v)}\\
  & \sum_{p=1}^\infty\frac{1}{p!}
  \int_\epsilon^{\rho_a} \prod_{i=1}^p \frac{d\rho_i}{\rho_i}R'(\rho_i)\,
  \rho\delta\Big(\rho-\rho_a-\sum_{i=1}^p \rho_i\Big)\,
  \Theta\Big(\frac{\rho_a}{\rho}>1-\tau\Big) \nonumber.
\end{align}
This expression is a little more complex than the corresponding one
for the cumulative distribution because the integration over $\rho_a$
can no longer trivially be done and the sum over the other $\rho_i$
now depends on $\rho_a$.
Nevertheless, we see the same two regimes appearing:
$\rho-\rho_a<\rho_a$ (\ie $\rho_a>\rho/2$) and $\rho-\rho_a>\rho_a$
(\ie $\rho_a<\rho/2$).
The former implies that each $\rho_{i\neq a} \approx \rho-\rho_a$
automatically satisfying $\rho_{i\neq a} < \rho-\rho_a$, while the
second implies that $\rho_i\approx \rho_a$.  As in the double
differential case, we rescale all the $\rho_i$ by $\rho-\rho_a$,
setting $R'(\rho_i)\approx R'(\rho-\rho_a)$, in the first case and by
$\rho_a$, setting $R'(\rho_i)\approx R'(\rho_a)$ in the second.
After some algebraic manipulation we find
\begin{align}
  H(\rho,<\tau)
  =
  \int_{(1-\tau)\rho}^\rho\frac{d\rho_a}{\rho_a}R'(\rho_a)
  &\Bigg[\frac{\rho}{\rho-\rho_a}R'(\rho-\rho_a)\frac{e^{-R(\rho-\rho_a)-\gamma_ER'(\rho-\rho_a)}}{\Gamma(1+R'(\rho-\rho_a))}\Theta(\rho_a>\rho/2)\\
  &+\frac{\rho}{\rho_a}R'(\rho_a)\frac{e^{-R(\rho_a)-\gamma_ER'(\rho_a)}}{\Gamma(1+R'(\rho_a))}f_\text{ME}\Big(\frac{\rho-\rho_a}{\rho_a};R'(\rho_a)\Big)\Theta(\rho_a<\rho/2)\Bigg].\nonumber
\end{align}
Note that the second line, where $\rho_a<\rho/2$, only contributes for
$\tau>\tfrac{1}{2}$.
We proceed by making the following simplification:
\[
R(\rho_a) \approx R(\rho) + R'(\rho)
\log\Big(\frac{\rho}{\rho_a}\Big),
\qquad\text{ and }\qquad
R'(\rho_a) \approx R'(\rho),
\]
in the second line, valid within our accuracy.
Correspondingly, for the first line, we expand $R'(\rho-\rho_a)$
around $\rho\tfrac{\tau}{1-\tau}$ so to avoid introducing a
discontinuity at $\tau=\tfrac{1}{2}$,\footnote{Note that, in this
  paper, we are not interested the limit $\tau\to 1$ which is
  definitely outside the phenomenologically-interesting region. This
  would require an additional resummation of logarithms of
  $1-\tau$. Practically, this would also mean exploring the region
  where a large number of emission significantly contribute to $\rho$,
  which would probably require to go beyond the approximation
  $R'(\zeta_i(1-\tau)\rho)\approx R'((1-\tau)\rho)$.} \ie
\[
  R(\rho-\rho_a) \approx R\Big(\rho\frac{\tau}{1-\tau}\Big)
  + R'\Big(\rho\frac{\tau}{1-\tau}\Big)
\log\Big(\frac{\rho\tau}{(1-\tau)(\rho-\rho_a)}\Big),
\quad\text{ and }\quad
R'(\rho-\rho_a) \approx R'\Big(\rho\frac{\tau}{1-\tau}\Big).
\]
Finally, for the emission ``$\rho_a$'', we replace the
$R'(\rho_a)$ factor in front of the square bracket by $R'(\rho)$.
There is obviously some arbitrariness in choosing the scale for all
these expansions (see also the discussion at the end of
Section~\ref{sec:diff-xsect}). We have checked explicitly that the
different choices are within the uncertainties described below.
Introducing $\tilde\tau=\tfrac{\tau}{1-\tau}$, 
we can write $H(\rho,<\tau)$ as
\begin{align}
  H(\rho,<\tau)
  =
  \int_{(1-\tau)\rho}^\rho\frac{d\rho_a}{\rho_a}R'(\rho)
  &\Bigg[R'(\rho\tilde\tau)\Big(\frac{\rho-\rho_a}{\rho}\Big)^{R'(\rho\tilde\tau)-1}\Big(\frac{1-\tau}{\tau}\Big)^{R'(\rho\tilde\tau)}\frac{e^{-R(\rho\tilde\tau)-\gamma_ER'(\rho\tilde\tau)}}{\Gamma(1+R'(\rho\tilde\tau))}\Theta(\rho_a>\rho/2)\nonumber\\
  &+R'(\rho)\Big(\frac{\rho_a}{\rho}\Big)^{R'(\rho)-1}\frac{e^{-R(\rho)-\gamma_ER'(\rho)}}{\Gamma(1+R'(\rho))}f_\text{ME}\Big(\frac{\rho-\rho_a}{\rho_a};R'(\rho)\Big)\Theta(\rho_a<\rho/2)\Bigg].\nonumber
\end{align}

The $\rho_a$ integration can only be done explicitly for
$\rho_a>\rho/2$, which gives
\begin{align}
  H(\rho,<\tau)
  & \overset{\tau<1/2}=
  R'(\rho)\frac{e^{-R(\rho\tilde\tau)-\gamma_ER'(\rho\tilde\tau)}}{\Gamma(1+R'(\rho\tilde\tau))}(1-\tau)^{R'(\rho\tilde\tau)}{}_2F_1(1,R'(\rho\tilde\tau);1+R'(\rho\tilde\tau);\tau)\label{eq:tau21-cumulative}\\
  & \overset{\tau>1/2}=
    R'(\rho)\frac{e^{-R(\rho)-\gamma_ER'(\rho)}}{\Gamma(1+R'(\rho))}\Big[2^{-R'(\rho)}{}_2F_1(1,R'(\rho);1+R'(\rho);\tfrac{1}{2})+R'(\rho)
    {\cal I}_\text{ME}\Big(\frac{\tau}{1-\tau};R'(\rho)\Big)\Big],\nonumber
\end{align}
with $_{2}F_1$ the Gauss hypergeometric function and
\[
  {\cal I}_\text{ME}(x;R')=\int_1^x \frac{du}{(1+u)^{R'}} f_\text{ME}(u;R').
\]
Eq.~(\ref{eq:tau21-cumulative}) is the main result of this paper.
We note that, at least for $\tau<\tfrac{1}{2}$, one mostly
recovers a simple resummed result, with finite $\tau$ effects present under the form of a
hypergeometric factor.

As for the case of the double-differential distribution, the above
expression does not take into account the effect of secondary
emissions. These contribute only when $\tau<\tfrac{1}{2}$ and can be
inserted by undoing the $z_a$ integration that leads to the
overall factor $R'(\rho)$:\footnote{These expression only differ from those used for the double-differential calculation by subleading
  factors of $1-\tau$.}
\begin{equation}
  R'(\rho) = \int_\rho^1 d z_a\,P(z_a)\,\frac{\alpha_s(\sqrt{z_a\rho}\,p_tR)}{2\pi}.
\end{equation}
and redefining $R$ and $R'(\rho\tilde\tau)$:
\begin{align}
  R(\rho\tilde\tau;z_a) & = \int_0^1\frac{d\theta^2}{\theta^2}\,dz\,P(z)
    \frac{\alpha_s(z\theta p_tR)}{2\pi}\,
    \Theta(z\theta^2>\rho\tilde\tau)\nonumber\\
   & + \int_0^{\theta_1^2}\frac{d\theta_{12}^2}{\theta_{12}^2}\int_0^1\,dz\,P(z)
    \frac{\alpha_s(zz_a\theta_{12} p_tR)}{2\pi}\,
     \Theta\Big(z\Big(\frac{\theta_{12}}{\theta_a}\Big)^2>\tilde\tau\Big).
     \label{eq:Rrhotautilde}
\end{align}

In the rest of the paper we focus on studying the effect of
the $\tau_{21}<\tau$ cut itself.  For this reason,
we define the normalised cumulative distribution
\begin{equation}\label{eq:Hnorm}
H_{\text{norm}}(\rho,<\tau) = \frac{H(\rho,<\tau)}{H(\rho)},
\end{equation}
where we use\footnote{Both $H(\rho,<\tau)$ and $H(\rho)$ neglect
  single-logarithmic contributions from soft-and-large-angle gluon
  radiation, including non-global logarithms. Although they would have
  to be included in a full NLL description of $H(\rho,<\tau)$, they
  can be neglected when it comes to discussing the effects of a cut on
  $\tau$. We will see in the next section that they can also be
  avoided altogether at our accuracy by working with groomed jets.}
\[
H(\rho) = R'(\rho)\frac{e^{-R(\rho)-\gamma_ER'(\rho)}}{\Gamma(1+R'(\rho))}.
\]

\subsection{$N$-subjettiness for a SoftDropped jet}\label{sec:tau21-sd}

Practical applications of jet substructure techniques almost always
use a groomed jet mass instead of the plain jet mass.
In this section, we discuss how our results can be adapted to the case
where both the mass and $\tau_{21}$ are calculated on a jet groomed
with the SoftDrop procedure.

The calculation done earlier in Sec.~\ref{sec:calc-finite-tau} can be
applied to the case of SoftDropped jets by replacing the radiators for
the plain jet (and their derivatives) by their SoftDrop counterparts.
One however has to be careful with our definition of these
objects: since the SoftDrop procedure stops its Cambridge/Aachen
declustering once it has found two subjets satisfying the SoftDrop
criterion, $z>z_{\text{cut}}\theta^\beta$, all emissions at smaller
angles are kept, whether or not they satisfy the SoftDrop
criterion, as already seen in Eq.~(\ref{eq:Rtau_SD}).
This suggests that in order to define the SoftDrop radiator,
$R_{\text{SD}}$, we need to isolate the largest-angle emission that
passes the SoftDrop condition.
The key result is that, at our accuracy, we can use the emission
that dominates the (SoftDrop) mass.
To see this, consider the situation where we have an emission, say
$a$, which dominates the (SoftDrop) mass, together with another
emission, say $b$, at larger angle and smaller mass passing the
SoftDrop condition.
At some mass scale $\rho_0$, one then defines the radiator with the
constraints
\begin{equation}\label{eq:sd-conditions-angle}
  \Theta(z>\zcut\theta^\beta\text{ or }\theta<\theta_b)\,
  \Theta(z\theta^2>\rho_0).
\end{equation}
We want to show that we can replace $\theta_b$ by $\theta_a$ in the
above constraint and forget about emission $b$. 
According to our above calculation, we need $R_\text{SD}$ (and
$R'_\text{SD})$ down to a scale $\rho_0$, typically $\rho\tau$ or
$\rho(1-\tau)$, which is at least as large as the second most massive
emission in the jet (see for example Eq.~(\ref{eq:tau21})).
This scale is always at least $\rho_b$. Since emission $b$ passes the
SoftDrop condition, the mass constrain in
(\ref{eq:sd-conditions-angle}) implies $z\theta^2>\rho_b$.
Using this and the fact that emission $b$ passes the SoftDrop
condition, we can easily see that the SoftDrop constraint
in~(\ref{eq:sd-conditions-angle}) is fully given by
$z>\zcut\theta^\beta$, and hence can be replaced by the condition
``$z>\zcut\theta^\beta\text{ or }\theta<\theta_a$'', since
$\theta_b>\theta_a$.
Obviously, in the complementary case where one emission, $a$, is both
the largest-mass and largest-angle emission passing the SoftDrop
condition, the constraint~(\ref{eq:sd-conditions-angle}) trivially has
$\theta_b$ replaced by $\theta_a$.
Note that since SoftDrop would stop at most when declustering
emission $a$, secondary emissions remain exactly as for the case of
the plain jet mass.

In conclusion this means that the calculation of the cumulative
distribution $H_\text{SD}(\rho,>\tau)$ for SoftDrop jets, proceeds in the 
same fashion as that presented in Sec.~\ref{sec:cumul-distrib}, up to a
redefinition of the radiators (using $\theta_a^2=\rho/z_a$):
\begin{align}
  R_\text{SD}(\rho) & = \int_0^1\frac{d\theta^2}{\theta^2}\,dz\,P(z)
    \frac{\alpha_s(z\theta p_tR)}{2\pi}\,
    \Theta(z>\zcut\theta^\beta)\,
    \Theta(z\theta^2>\rho\tilde\tau)\\
  R_\text{SD}(\rho\tilde\tau;z_a) & = \int_0^1\frac{d\theta^2}{\theta^2}\,dz\,P(z)
    \frac{\alpha_s(z\theta p_tR)}{2\pi}\,
    \Theta(z>\zcut\theta^\beta\text{ or }\theta>\theta_a)\,
    \Theta(z\theta^2>\rho\tilde\tau)\nonumber\\
   & + \int_0^{\theta_a^2}\frac{d\theta_{12}^2}{\theta_{12}^2}\int_0^1\,dz\,P(z)
    \frac{\alpha_s(zz_a\theta_{12} p_tR)}{2\pi}\,
     \Theta\Big(z\Big(\frac{\theta_{12}}{\theta_a}\Big)^2>\tilde\tau\Big),
     \label{eq:Rrhotautildesd}
\end{align}
and correspondingly for $R'_\text{SD}$.
Additionally, when integrating over $z_a$, the lower bound of
integration should be set to the lowest value allowed by the SoftDrop
condition, i.e.
\[
  z_a > (\zcut^2\rho^\beta)^{\frac{1}{2+\beta}}.
\]

\subsection{Scale uncertainties and matching to fixed order}\label{sec:uncertainties}

Given the discussion above about the freedom in setting the scale
entering the radiators while keeping the same formal accuracy, it is
interesting to consider adding a scale uncertainty to our results.
Here, we consider two possible source of uncertainty: the
renormalisation and resummation scale uncertainties. The former is
accounted for by varying the ``hard scale'', $p_tR$, at
which we compute the coupling by a factor $\mu_R=1/2$ or $2$.
To assess the resummation scale uncertainty, we vary the
reference scale $p_tR$ in the definition of the logarithm of $\rho$ by
a factor $\mu_Q=1/2$ or $2$. Since our calculation includes
single-logarithmic terms in $\rho$, we need to introduce an extra
contribution to the exponentials in Eq.~\eqref{eq:tau21-cumulative} to
correct for the single-logarithmic term generated by the
double-logarithmic radiator $R(\rho\tau)$. For
$\rho=\mu_Q\tfrac{m^2}{(p_tR)^2}$, we make the replacement
\begin{equation}\label{eq:muQrescaling}
R(\rho\tau) \to R(\rho\tau)+R'(\rho\tau)\log(\mu_Q),
\end{equation}
and a similar expression for $R(\rho)$.
Our final uncertainty is taken as the envelope of the $\mu_R$ and
$\mu_Q$ variations.

\section{Comparison to Monte Carlo simulations}\label{sec:mc-tests}

\subsection{Results at fixed order}\label{sec:mc-tests-fo}

\begin{figure}[!t]
  \centering
  \includegraphics[width=0.48\textwidth,page=1]{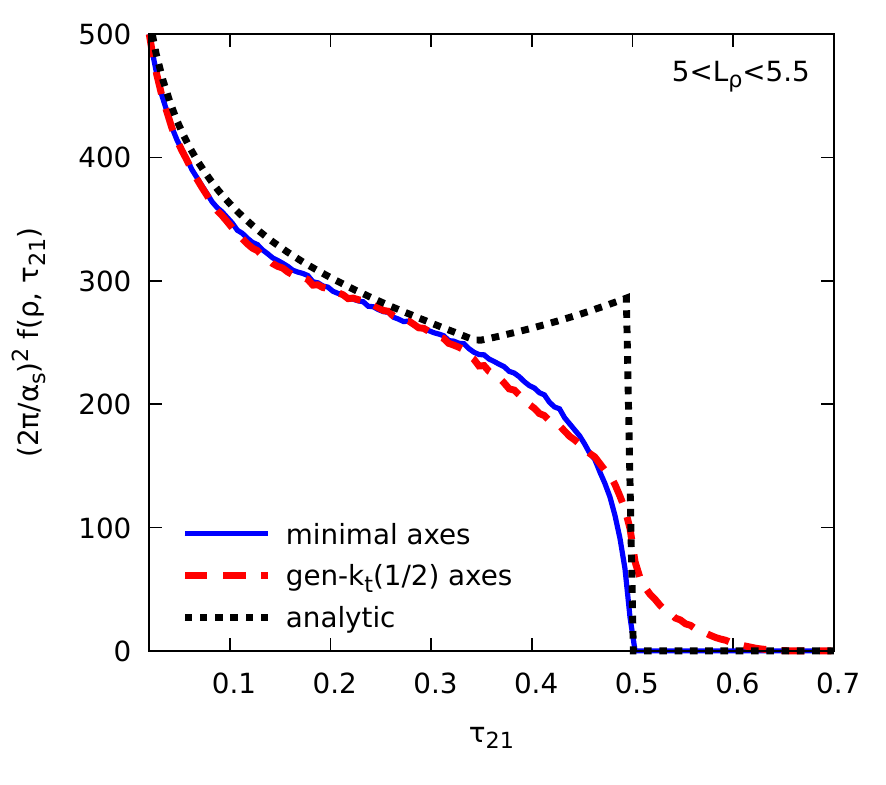}%
  \hfill
  \includegraphics[width=0.48\textwidth,page=2]{figs/tau21-distrib-event2.pdf}%
  \caption{Comparison of our analytic results (dotted black) with the Event2
    generator for the $\tau_{21}$ distribution in a bin of $\rho$.
    For the Event2 simulations, we show results for both the
    generalised-$k_t$ (solid red) and minimal axes (dashed blue).
    The left and right plots corresponds to different bins in $L_\rho=\log(1/\rho)$. 
  }\label{fig:event2-tau21-distrib}
\end{figure}

We first compare our results with a fixed-order calculation
at the first order where a non-trivial $\tau$ dependence appears:
${\cal O}(\alpha_s^2)$.
In this case, we consider the distribution
\begin{equation}\label{eq:hgreater}
  H(\rho,>\tau) = H(\rho) - H(\rho,<\tau),
\end{equation}
for a jet to have a given ``mass'' $\rho$ and a $\tau_{21}$ ratio be
above a cut $\tau$.
We do this as in this case we are sensitive to the situation
with two real emission in the jet.
To compute in our approximation Eq.~\ref{eq:hgreater}, we
expand Eq.~(\ref{eq:tau21-cumulative}) to order $\alpha_s^2$, which gives
\begin{equation}
H(\rho,>\tau) = R'(\rho)\, [ R(\rho \tilde\tau)  - R(\rho) ]\,\Theta\big(\tau<\tfrac{1}{2}\big).
\end{equation}
We can then proceed by expanding $R$ and $R'$ in $\alpha_s$
(equivalent to using a fixed-order prescription), which gives the leading
logarithmic contribution in $\rho$.
\begin{align}\label{eq:cum-fixed-order}
  H_\text{plain}(\rho,>\tau) &= \Big(\frac{\alpha_sC_F}{\pi}\Big)^2
                               \log\Big(\frac{1-\tau}{\tau}\Big)\log^2(\rho)
                               + {\cal O}(\log(\rho)),\\
  H_\text{SD}(\rho,>\tau)    &=  \Big(\frac{\beta}{2+\beta}\frac{\alpha_sC_F}{\pi}\Big)^2
                               \log\Big(\frac{1-\tau}{\tau}\Big)\log^2(\rho)
                               + {\cal O}(\log(\rho)),
\end{align}
respectively for the plain jet and for a SoftDropped jet.
Additionally, we take the derivative of Eq.~(\ref{eq:tau21-cumulative})
with respect to $\tau_{21}$,
and perform a similar expansion, yielding the leading $\log(\rho)$ 
contribution as well as the double logarithmic term in $\tau_{21}$
\begin{align}\label{eq:diff-fixed-order}
  f_\text{plain}(\rho,\tau_{21})
  = \Big(\frac{\alpha_sC_F}{\pi}\Big)^2
  &\frac{1}{1-\tau}
  \Big(\log\Big(\frac{1}{\rho}\Big)+B_q\Big)\\
  & \times \bigg[
  \Big(\log\Big(\frac{1}{\rho}\Big)+\log\Big(\frac{1-\tau}{\tau}\Big)+B_q\Big)
  +\frac{C_A}{C_F}\Big(\log\Big(\frac{1-\tau}{\tau}\Big)+B_g\Big)\bigg].\nonumber
\end{align}
Note that we do not include in this case the similarly derived expressions
in the SoftDrop jet case, as they are
more lengthy and complex due to the constraint (\ref{eq:sd-conditions-angle}.

To check our accuracy claim, we test to what extent our
approximations, Eqs~(\ref{eq:cum-fixed-order}) and (\ref{eq:diff-fixed-order}),
can reproduce a fixed-order prediction for the same observable.
For this purpose, we use the
Event2~\cite{Catani:1996jh,Catani:1996vz} generator, by rotating events to
align them along one of the axis and proceed as if they were $pp$
collisions.

Additionally, to simulate the leading behaviour we are interested in,
we compute the same quantities by integrating the triple collinear
splitting function~\cite{Campbell:1997hg,Catani:1998nv,Catani:1999ss}
without imposing any ordering or soft approximation.  In small-$R$
limit this has the same $\alpha_s^2\log^2(\rho)$ and
$\alpha_s^2\log(\rho)$ logarithmic dependence as an exact fixed-order
calculation, for the plain jet (see also Sec.~3.2
of~\cite{Dasgupta:2018emf}).
This validity extends to larger $R$ in the SoftDrop case, which keeps
only emissions at angles that are suppressed by powers of $\rho$. The
high level of agreement between the two can be seen in the results
presented in this section.  We present this approximation too as it
can be more easily pushed numerically to smaller values of $\rho$ and
$\tau$, which one is interested in, and it might be easier to use in
the case of matching to fixed-order.

We start by showing the $\tau_{21}$ distribution in
Fig.~\ref{fig:event2-tau21-distrib}.
At small $\tau_{21}$ ($\tau_{21}\lesssim 0.3$) our analytic
results are in perfect agreement with the exact Event2 results.
For larger $\tau_{21}$, we see a transition at
$\tau_{21}=\tfrac{1}{2}$, as already discussed in
Sec.~\ref{sec:tau-value}. Although this transition is present in the 
Event2 simulations as well, it appears to be smoother.
Around the transition point, we also see some differences between the
two choices of axes as well as a shoulder in the analytic calculation
which is absent in the Event2 simulations. Increasing $\log(1/\rho)$,
makes the transition at $\tau_{21}=\tfrac{1}{2}$ in Event 2
become sharper. This is expected, as for large $\log(1/\rho)$ our calculation 
captures the dominant $\alpha_s^2\log^2(1/\rho)$ contribution to
$f(\rho,\tau)$, leaving corrections of order
$\alpha_s^2\log(1/\rho)$.
However this does not obviously seem the case in the shoulder region
where the difference between Event2 and our analytic results seems to
increase as rapidly as the rest of the distribution.
We traced back this shoulder effect to differences between the exact
$\tau_{21}$ and our leading-logarithmic approximation,
Eq.~(\ref{eq:tau21}), specifically in the region of similar angles
($\theta_2\sim \theta_1$ in Fig.~\ref{fig:lund}).
We discuss this in more details in Appendix~\ref{app:subleading} where
we show that this region indeed only gives subleading corrections and
that these corrections are increasingly numerically relevant when
approaching $\tau_{21}=\tfrac{1}{2}$.

\begin{figure}[!t]
  \centering
  \includegraphics[width=0.48\textwidth,page=1]{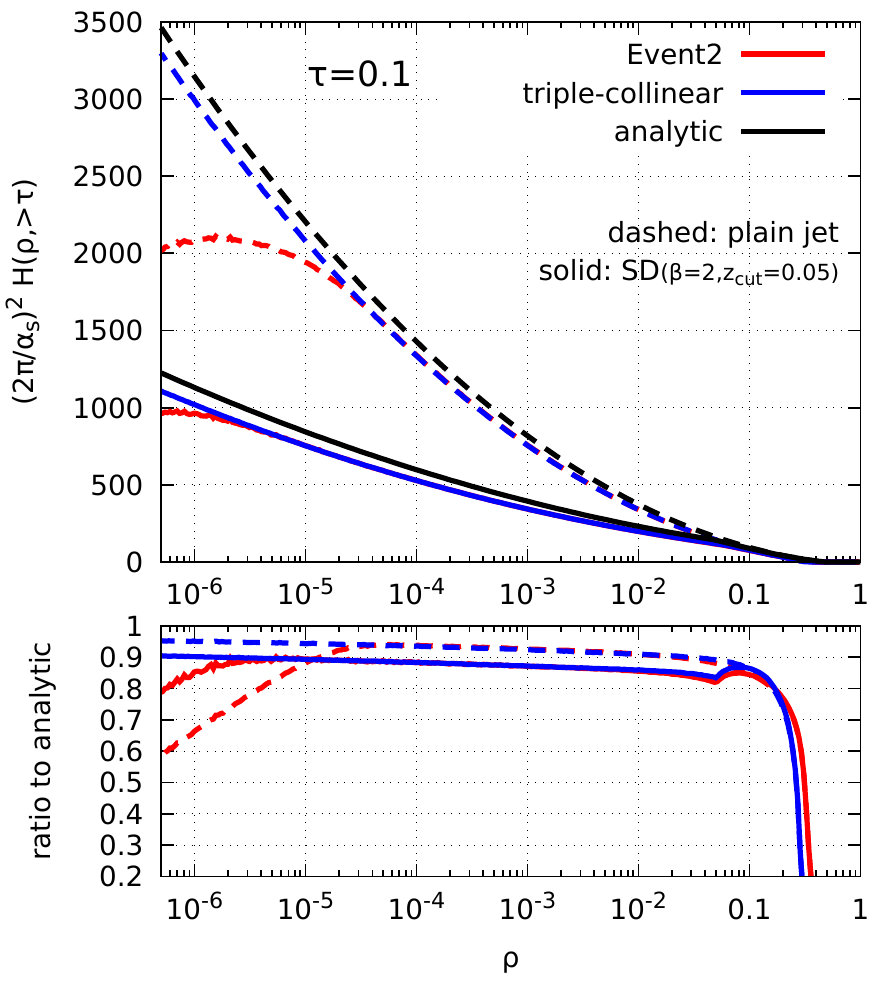}%
  \hfill
  \includegraphics[width=0.48\textwidth,page=3]{figs/event2-distrib.pdf}%
  \caption{Comparison of our analytic results with the Event2
    generator and with the triple-collinear splitting function for the
    $H(\rho,>\tau)$ distribution.
    Dashed and solid lines respectively correspond to results obtained
    using plain and SoftDropped jets.
    The top panel shows $H(\rho,>\tau)$ and the bottom panel
    shows the ratio to our analytic results.
    The left and right plots correspond to a cut $\tau=0.1$ and
    $\tau=0.3$, respectively.}\label{fig:event2-distrib}
\end{figure}

\begin{figure}[!t]
  \centering
  \includegraphics[width=0.48\textwidth]{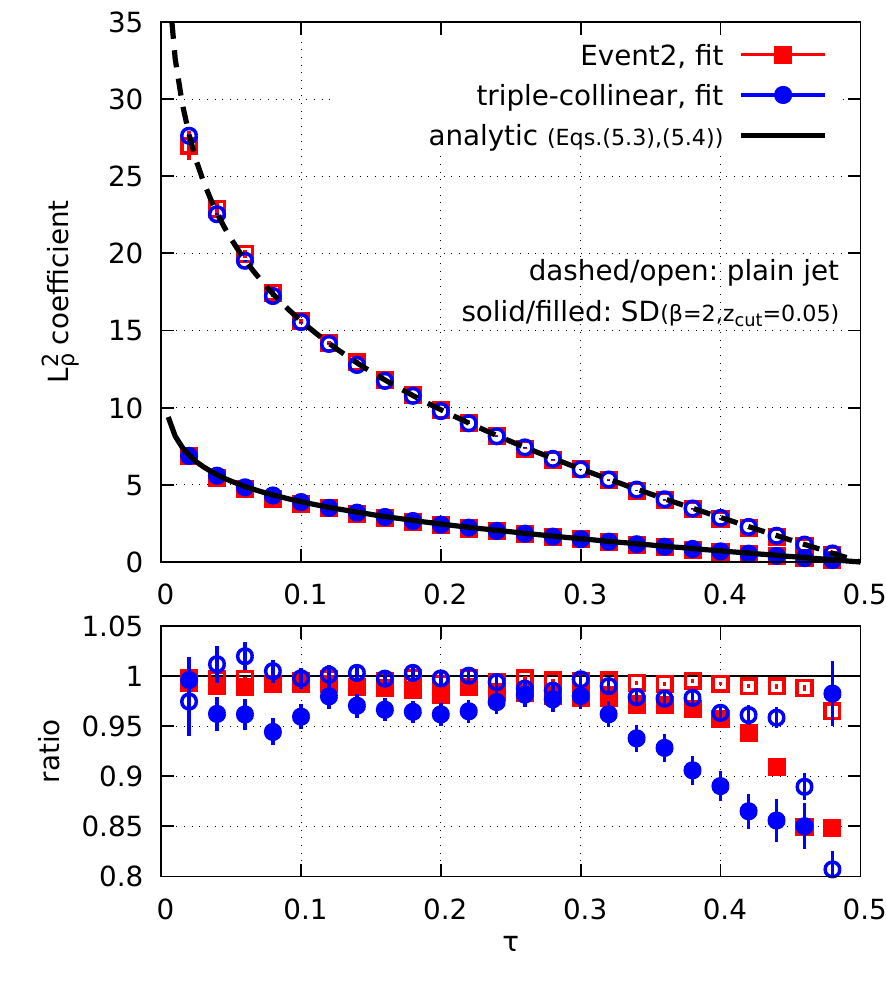}%
  \caption{Coefficient of the $\log^2(1/\rho)$ contribution to the
    $H(\rho,>\tau)$ distribution.
    For the Event2 results (filled symbols) and the results obtained by
    integrating over the triple-collinear splitting function (open
    symbols), we perform a fit to the numerical results for
    each value of $\tau$.
    (Red) circles correspond to results obtained from the plain jet,
    and (black) squares to results where SoftDrop has first been
    applied.
    The plot insert shows the ratio to the expected analytic result.
  }\label{fig:event2-Lrho2-coef}
\end{figure}

A comparison of the $H(\rho,>\tau)$ distribution, obtained either from
Event2 or integrating over the triple-collinear splitting function, to
our analytic results is presented on Fig.~\ref{fig:event2-distrib}. We
plot $H(\rho,>\tau)$ as a function of $\rho$ for two different
values of the $\tau$ cut. Results are shown for both the plain jet and
SoftDropped jet (using $\beta=2$ and $\zcut=0.05$).
As it can be seen, our calculation indeed captures the dominant
$\log^2(1/\rho)$ behaviour.\footnote{The deviations close to
  $\tau=\tfrac{1}{2}$ can be attributed to the shoulder in the
  $\tau_{21}$ distribution which slows down the convergence in that
  region.}
This is confirmed by Fig.~\ref{fig:event2-Lrho2-coef} which shows the
result for the coefficient of the leading $\log^2(1/\rho)$
contribution. For the Event2 and the triple-collinear results, we
extract this coefficient using a simple fit of the distribution, for
each individual cut on $\tau$. The fitted coefficient lies very closely
to the expected analytic results, for both the plain and SoftDropped
jets. We believe that the small discrepancy is related to the limited
fitting range and the difficulty to obtain numerical results in the
very small $\rho$ limit.

We also see in Fig.~\ref{fig:event2-distrib} that the triple-collinear
results are almost identical to what is obtained from Event2, except
in the large $\rho$ region where the triple-collinear approximation
breaks down, and in the small $\rho$ region, where Event2 has an
infrared cut-off causing the drop seen in the figure.

\subsection{Parton shower Monte Carlo simulations}\label{sec:v-parton-shower}

\begin{figure}[!t]
  \centering
  \includegraphics[width=0.48\textwidth,page=1]{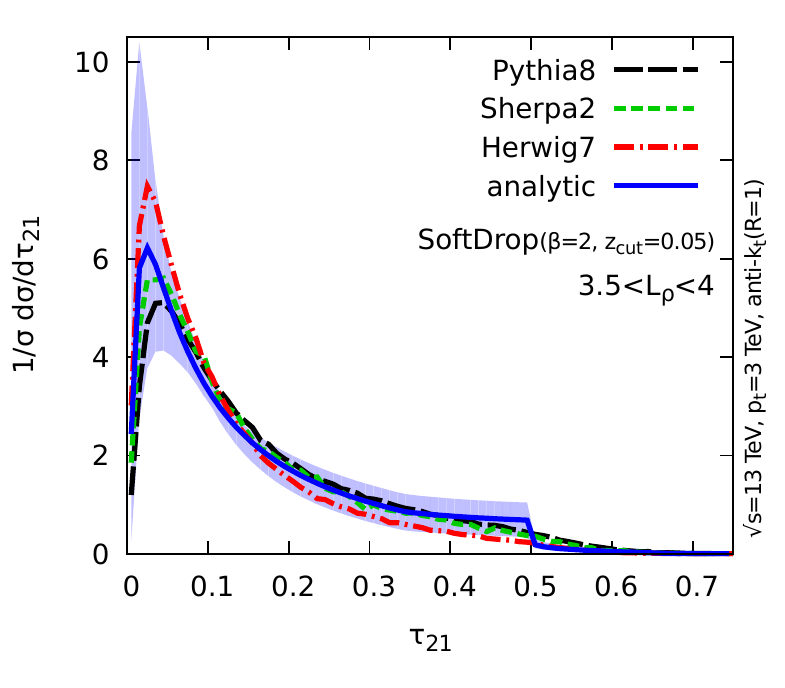}\hfill%
  \includegraphics[width=0.48\textwidth,page=4]{figs/tau21-distribs.pdf}\\
  \includegraphics[width=0.48\textwidth,page=7]{figs/tau21-distribs.pdf}\hfill
  \begin{minipage}[b]{0.42\textwidth}
    \caption{Differential distribution in $\tau$ compared to Monte Carlo
      for 3 different bins in $L_\rho=\log(1/\rho)$. We show our
      analytic results (with their uncertainty band) compared to the
      Pythia8, Sherpa2 and Herwig7 generators. All results are for jets
      groomed with SoftDrop. The vertical dashed lines indicate when the
      scale $\rho\tau$ starts to hit a given $k_t$ value, indicative of
      where non-perturbative effects are expected to become
      dominant.}\label{fig:tau-distrib}
    \vspace*{1.4cm}
  \end{minipage}\hspace*{0.3cm}
\end{figure}

\begin{figure}[!t]
  \centering
  \includegraphics[width=0.48\textwidth,page=1]{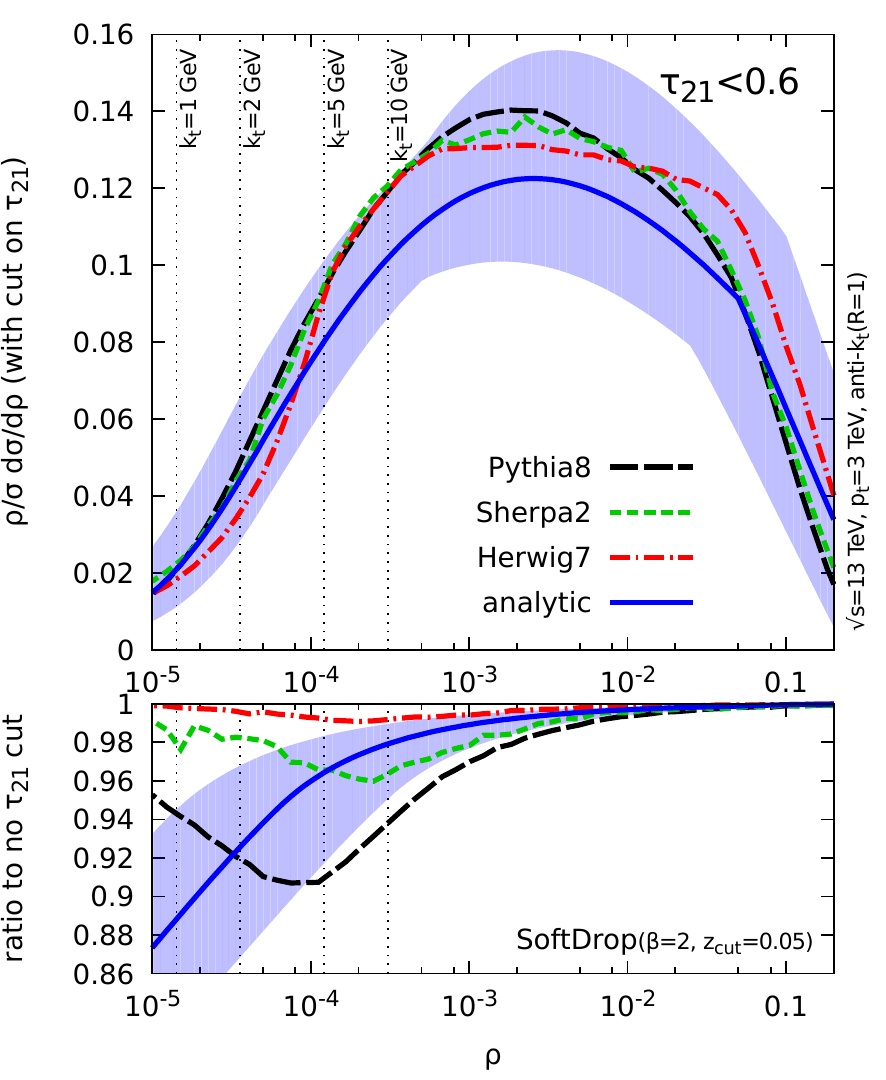}\hfill%
  \includegraphics[width=0.48\textwidth,page=4]{figs/rho-distribs-with-ratio.pdf}\\
  \includegraphics[width=0.48\textwidth,page=6]{figs/rho-distribs-with-ratio.pdf}\hfill
  \begin{minipage}[b]{0.4\textwidth}
    \caption{Mass distribution with a cut on $\tau_{21}$ compared to
      Monte Carlo for 3 different values of the cut.
      We show our analytic results (with their uncertainty band)
      compared to the Pythia8, Sherpa2 and Herwig7 generators. All
      results are for jets groomed with SoftDrop. The vertical dashed
      lines indicate when the scale $\rho\tau$ starts to hit a given
      $k_t$ value, indicative of where non-perturbative effects are
      expected to become dominant.}\label{fig:rho-distrib}
    \vspace*{2.3cm}
  \end{minipage}\hspace*{0.5cm}
\end{figure}

\paragraph{Setup.}
We now compare our analytic result to parton shower Monte Carlo
generators.
For this, we simulate dijet events with three different
generators: Pythia~8.230~\cite{Sjostrand:2014zea} (Monash13
tune~\cite{Skands:2014pea}), Sherpa~2.2.4~\cite{Gleisberg:2008ta} and
Herwig~7.1.1~\cite{Corcella:2002jc,Bellm:2017bvx} with angular-ordered
shower.
We only consider underlying fixed order matrix elements with quarks in the final
states, which means that  we can assume quark jets for our analytic results as well.
Events are simulated at $\sqrt{s}=13$~TeV and we focus for the
moment on parton level results.
We reconstruct jets with the anti-$k_t$~
algorithm~\cite{Cacciari:2008gp} with $R=1$ using
FastJet~3.3.1~\cite{Cacciari:2005hq,Cacciari:2011ma}.
We further require that all jets have $p_t>3$~TeV.
We apply SoftDrop, using $\beta=2$ and $\zcut=0.05$, to each jet and
compute the jet mass and $N$-subjettiness on the SoftDropped jet. For
$\tau_1$ and $\tau_2$ we use the generalised-$k_t$($p=1/2$)
(difference wrt to minimal axes in this case are smaller than what we
observe with Event2).
We then consider two distributions: either the $\tau_{21}$
distribution for jets within a restricted window of mass, or the jet
mass distribution for a given cut on $\tau_{21}$.
All analytical results shown here are obtained
from the cumulative distribution computed in
Sec.~\ref{sec:cumul-distrib} (by taking the $\tau$ derivative to get
the $\tau_{21}$ differential distribution), applied to SoftDropped
jets (see Sec.~\ref{sec:tau21-sd}), with the uncertainty band
calculated as described in Sec.~\ref{sec:uncertainties}. For the
radiators, we use the expressions reported in
Appendix~\ref{app:radiators}, including running-coupling effects.

\paragraph{Comparison at parton level.}
Our results for the $N$-subjettiness distribution are presented in
Fig.~\ref{fig:tau-distrib} for three different bins in
$L_\rho=\log(1/\rho)$.
Overall, we see a good agreement between the Monte Carlo simulations
and our approximation, already at relatively small values of $\log(1/\rho)$,
with Herwig lying at the edge of our uncertainty band. 
As discussed in the previous sections, we expect and observe a
transition at $\tau=\tfrac{1}{2}$ in the analytic calculation, which
is smeared in Monte Carlo simulations. This can be explained by the fact they
compute the value of $\tau_{21}$ exactly.
Going above $\tau=\tfrac{1}{2}$, we observe, from our analytic
calculation, a sizeable contribution due to multiple emissions.
The dashed vertical lines on Fig.~\ref{fig:tau-distrib} indicate where
our calculation becomes sensitive to a given $k_t$ scale (with the
soft scale of our calculation taken as the lowest $k_t$ accessible for
a mass scale of $\rho\tau$). As $k_t$ decreases, we expect
sizeable non-perturbative contributions and we discuss this further
in the following paragraph.

Fig.~\ref{fig:rho-distrib} shows the mass distribution obtained for
three different cuts on $\tau_{21}$. The top panels show the raw mass
distribution,while the bottom panels show the distribution normalised
by the uncut mass distribution, highlighting the effect of the
$N$-subjettiness cut itself.
As expected, putting a tighter cut on $N$-subjettiness reduces the
mass distribution. As before, we see a good agreement between our
calculation and the Monte Carlo simulations, at least in the
perturbative region. We also see differences between the three
generators of the order of our estimated theory uncertainty.

\begin{figure} [!t]
  \centering
  \includegraphics[width=0.48\textwidth,page=1]{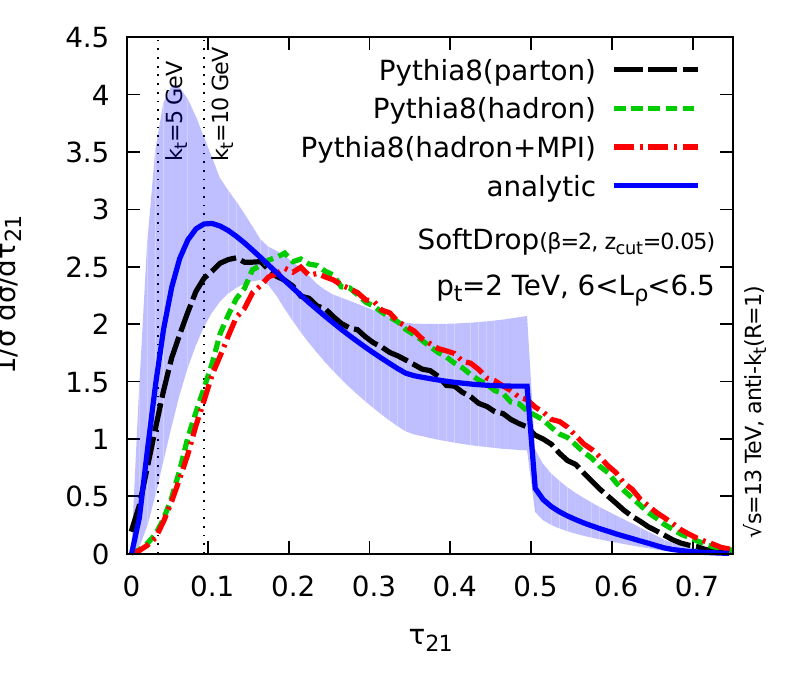}\hfill%
  \includegraphics[width=0.48\textwidth,page=2]{figs/tau21-distribs-np.pdf}\\
  \includegraphics[width=0.48\textwidth,page=3]{figs/tau21-distribs-np.pdf}\hfill
  \begin{minipage}[b]{0.4\textwidth}
    \caption{Same as Fig.~\ref{fig:tau-distrib} now for different values
      of $p_t$. Here we show results for the Pythia8 generator at
      different levels in order to gauge the importance of
      non-perturbative effects. For each $p_t$, the bin in $L_\rho=\log(1/\rho)$ is
      adjusted to be roughly around the mass of the $W$ boson (for
      definiteness).}\label{fig:tau-distrib-np}
    \vspace*{2.2cm}
  \end{minipage}\hspace*{0.5cm}
\end{figure}

\begin{figure}[!t]
  \centering
  \includegraphics[width=0.48\textwidth,page=1]{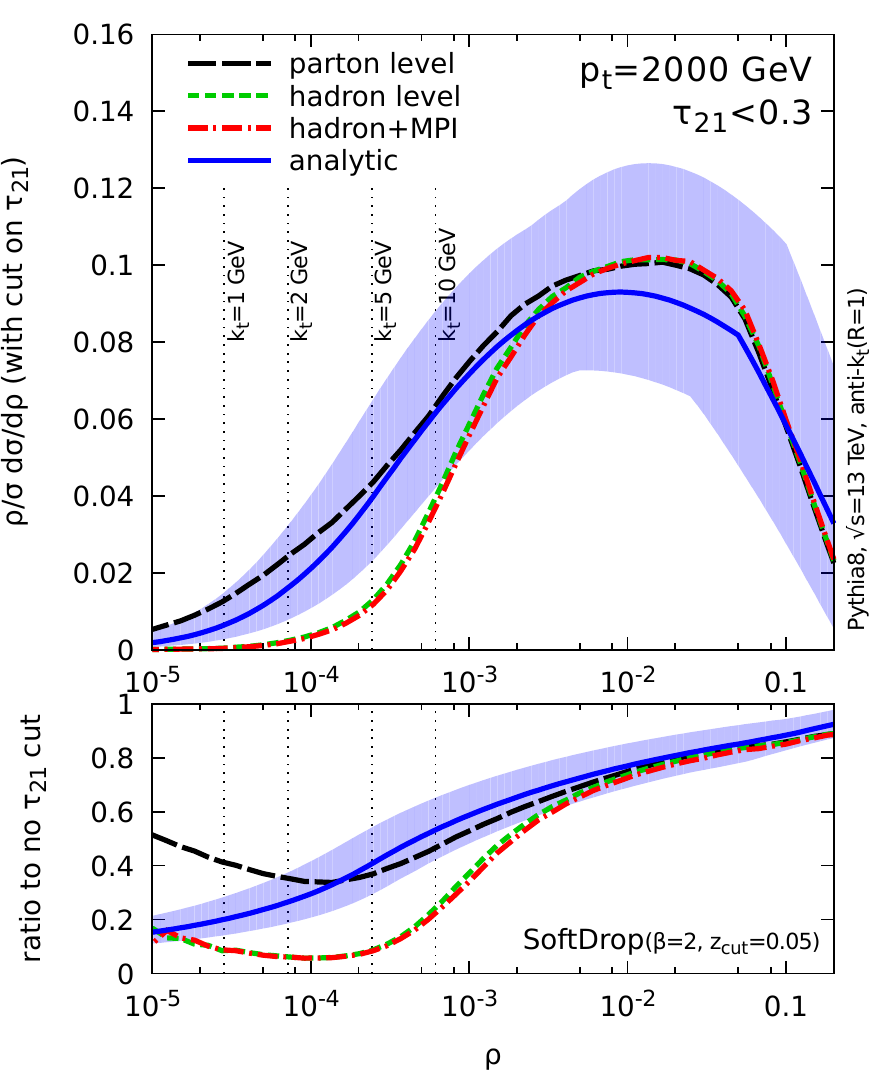}\hfill%
  \includegraphics[width=0.48\textwidth,page=2]{figs/rho-distribs-with-ratio-np.pdf}\\
  \includegraphics[width=0.48\textwidth,page=3]{figs/rho-distribs-with-ratio-np.pdf}\hfill
  \begin{minipage}[b]{0.4\textwidth}
    \caption{Same as Fig.~\ref{fig:rho-distrib} now for different values
      of $p_t$. Here we show results for the Pythia8 generator at
      different levels in order to gauge the importance of
      non-perturbative effects. All the plots use a representative  cut on $\tau_{21}$
      of 0.3.}\label{fig:rho-distrib-np}
    \vspace*{3.2cm}
  \end{minipage}\hspace*{0.5cm}
\end{figure}

\paragraph{Lower $p_t$ and non-perturbative effects.}
We now want to check the level of agreement of our prediction when the
jet $p_t$ is smaller and assess the importance of non-perturbative
corrections.
This is shown in Figs.~\ref{fig:tau-distrib-np}
and~\ref{fig:rho-distrib-np}, where the different plots correspond to
$p_t$ cuts of 2~TeV, 1~TeV and 500~GeV respectively. For each $p_t$ we
have adjusted the bin in $L_\rho=\log(1/\rho)$ to be roughly around
the value of the $W$ mass, a typical scale where the $\tau_{21}$ ratio
is used in phenomenological applications.
We show in these plots Pythia distributions obtained from different
type of events: parton level (long-dashed black lines), and hadron
level with both multiple-parton-interactions (MPI) switched off
(short-dashed green lines) and with MPI switch on (dash-dotted red
lines).
As far as the perturbative aspects are concerned, the agreement
between our calculation and Pythia remains valid for smaller boosts.
We see that hadronisation corrections have a sizeable
impact on the distributions, even in regions of phase-space, where
we are only sensitive to fairly large $k_t$ scales.
Furthermore, while MPI effects are small for 1 and 2~TeV jets,
they are sizeable for 500~GeV jets.
These effects can be reduced by using a more aggressive grooming
procedure, like a smaller value of $\beta$, \eg using the modified
MassDrop tagger (mMDT), or a larger value of $\zcut$.
In that context, note that we have checked that our analytic
calculations still work in the case of the mMDT where logarithms
of $\rho$ resummed in our multiple-emission contributions (the
$R'(\rho\tau)$ factors) are now replaced by logarithms of $\zcut$.

\begin{figure}[!t]
  \centering
  \includegraphics[width=0.48\textwidth,page=2]{figs/ddt.pdf}\hfill%
  \includegraphics[width=0.48\textwidth,page=1]{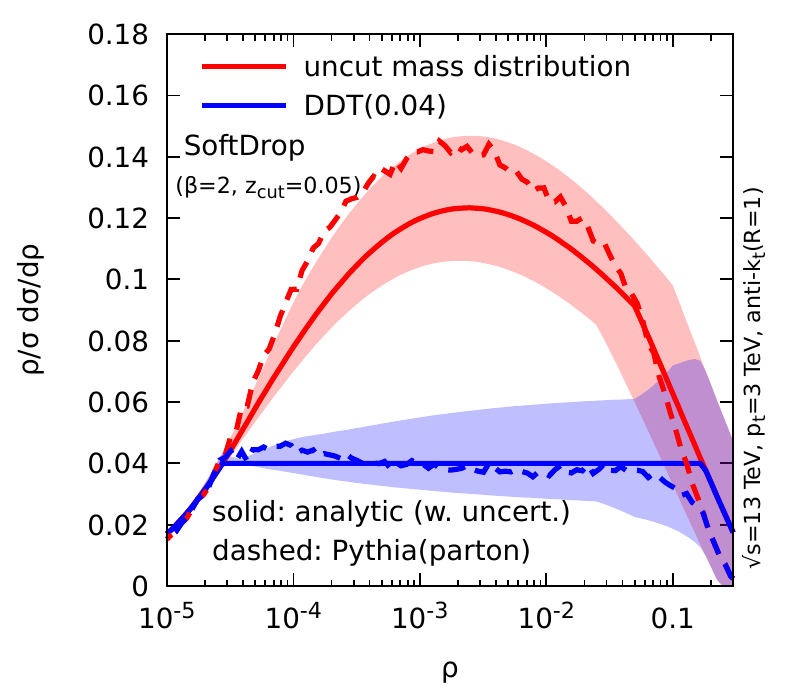}
  \caption{Analytic construction of a decorrelated tagger (DDT). Left:
    cut on $\tau_{21}$ which would give an analytic mass spectrum
    $\rho/\sigma d\sigma/d\rho=0.04$. Right: resulting mass spectrum
    analytically (with an uncertainty band) and using
    Pythia8.}\label{fig:ddt}
\end{figure}

\paragraph{Decorrelated taggers.}
One interesting application of our analytic control of
$N$-subjettiness cut is that it largely facilitates the design of a
decorrelated tagger~\cite{Dolen:2016kst}: for each value of the mass,
one can determine, based on our calculation, the value of the $\tau$
cut required to get a flat mass distribution at a given level, say,
with $\rho/\sigma d\sigma/d\rho$ somewhere in the 0.03-0.04 range
(lower values would start having a larger sensitivity to
non-perturbative effects).
We present the result of such a study in Fig.~\ref{fig:ddt}. For each
value of $\rho$, we adjust the cut on $\tau_{21}$ so as to
obtain $\rho/\sigma d\sigma/d\rho=0.04$. The cut one obtains is shown
in the left plot (whenever the uncut distribution was already smaller
than 0.04, we did not impose a further constraint on $\tau_{21}$).
The resulting distribution is shown in the right plot together with an
uncertainty band and the result of applying the same $\rho$-dependent
$\tau_{21}$ cut on a (parton-level) Pythia simulation.
We see that the resulting decorrelated distribution (labelled ``DDT'')
on the plot, in the Pythia simulation is almost flat, and at least
within our analytic uncertainty.
From a further study, one could conceive making a combined adjustment
of the $\tau$ cut together with the SoftDrop parameters in order to
obtain a flat background and maximise the signal efficiency for a
colourless 2-body decay like in the case of electroweak ($H/W/Z$) bosons.

\section{Conclusions}\label{sec:conclusions}

In perturbative QCD, boosted jets are characterised by large
logarithms of $m/p_t$, \ie the ratio of their mass to their transverse
momentum.
In this work we have shown how one can achieve an all-order
resummation of the dominant logarithms of the jet mass in the presence
of a cut on a jet shape. Compared to our previous work, we lift
the assumption that the cut is small.
This, in practice, allows one to take cut values of physical
relevance.
In this paper, we have focused on applying a cut on a particular jet
shape, namely the $N$-subjettiness $\tau_{21}$ ratio with the angular
exponent $\beta$ set to 2.
We compute both the $\tau_{21}$ distribution for a boosted jet,
and the jet mass distribution in the presence of a cut on $\tau_{21}$.
The calculation is structured so as to also include the leading
logarithms of the jet shape when it becomes small, hence recovering
results from previous works.

Besides the analytic results presented throughout the paper for
$\tau_{21}$, we are confident that the method can be applied to a wide
range of other jet shapes.
In a nutshell, the calculation is organised in a number of key steps:
(i) starting from a generic sum over any number of real emissions,
isolate the emission the dominates the jet mass, (ii) use the
shape to deduce the relevant physical scale for the remaining
emissions, (iii) simplify the expressions using CAESAR-like
techniques, standard in resummation calculations.
For more complex observables, one likely also have to isolate other
dominant emissions in step (i), like the emission dominating
the plain jet mass (potentially different from the one dominating the
groomed jet mass) in the case of a dichroic $N$-subjettiness ratio, or
the emission dominating the jet broadening (potentially different from
the one dominating the jet mass) in the case of the $\beta=1$
$\tau_{21}$ ratio. The generic approach presented here is then
expected to still apply.
In the future, we plan to explore other jet shapes like $\tau_{21}$
ratios for a generic $\beta$, dichroic ratios~\cite{Salam:2016yht} and
energy-correlation functions~\cite{Larkoski:2013eya}, as well as
investigating shapes relevant for (3-prong) top tagging like the
$\tau_{32}$ ratio.
Concerning energy correlation functions, it would be interesting to
compare our findings with results obtained in SCET e.g. for
$D_2$~\cite{Larkoski:2015kga,Larkoski:2017cqq,Moult:2017okx},
especially since $D_2$ appears to yield an efficient tagger (see
\eg~\cite{Bendavid:2018nar}).

We have compared our analytic predictions to the three most used Monte
Carlo event generators, Pythia, Herwig and Sherpa, in two cases: the
$\tau_{21}$ distribution and the jet mass distribution with a
$\tau_{21}$ cut. We have concentrated on the case of jets previously
groomed with SoftDrop, to limit non-perturbative effects. In both
cases, we see a good agreement with Monte Carlo predictions, within
our theoretical uncertainty band, in the region where resummation
matters.
As another example of a phenomenologically-relevant application of our
results, we have used our analytic calculations to build a
decorrelated tagger.

This work opens on several possible future developments. First, one
could try to extend the precision of our calculation to include
subleading logarithms and match it with fixed-order results.
(Note however that reaching an NLO accuracy for the fixed-order part of
the calculation would require $2\to 4$ QCD events at NLO.)
Such a prediction could then be compared to an
experimental measurement, similarly to what has been done recently for
the groomed jet
mass~\cite{CMS:2017tdn,Marzani:2017kqd,Marzani:2017mva,Aaboud:2017qwh}.
Finally, the theoretical uncertainty on our calculations, complemented
with an assessment of the non-perturbative uncertainties, could then
be used to estimate the theoretical uncertainty of boosted taggers
used in searches.

\section*{Acknowledgements}
GS and DN are supported by the French Agence Nationale de la
Recherche, under grant ANR-15-CE31-0016.
We wish to thank Gavin Salam for collaboration in the early stages of
this work, helpful discussions and comments on the manuscript.
We also thank Lais Schunk and Mrinal Dasgupta for discussions at
various stages of this project.

\appendix

\section{Explicit results for the radiators}\label{app:radiators}

The full expressions for the radiators and their derivatives are
already available from the literature (see
\eg~\cite{Marzani:2017kqd,Dasgupta:2015lxh,Banfi:2004yd}). We
summarise them here for completeness.

The SoftDrop radiator can be written as (assuming $\rho<\zcut$)
\begin{align}
&  R_\text{SD}(\rho)=\\
  &= \frac{C_i}{2\pi\alpha_s\beta_0^2}\bigg\{
     \bigg[
    W(1-\lambda_B)
    -\frac{W(1-\lambda_c)}{1+\beta}-2W\Big(1-\frac{\lambda_\rho+\lambda_B}{2}\Big)
    +\frac{2+\beta}{1+\beta}W\Big(1-\frac{\lambda_c+(1+\beta)\lambda_\rho}{2+\beta}\Big)
     \bigg]\nonumber\\
&
  +\frac{\alpha_s \beta_1}{\beta_0} \bigg[
  V(1-\lambda_B)
  -\frac{V(1-\lambda_c)}{1+\beta}
  -2V\Big(1-\frac{\lambda_\rho+\lambda_B}{2}\Big)
  +\frac{2+\beta}{1+\beta}V\Big(1-\frac{\lambda_c+(1+\beta)\lambda_\rho}{2+\beta}\Big)  
  \bigg]\nonumber\\
&
 -\frac{\alpha_s K}{2\pi}\bigg[
  \log(1-\lambda_B)
  -\frac{\log(1-\lambda_c)}{1+\beta}
  -2\log\Big(1-\frac{\lambda_\rho+\lambda_B}{2}\Big)
  +\frac{2+\beta}{1+\beta}\log\Big(1-\frac{\lambda_c+(1+\beta)\lambda_\rho}{2+\beta}\Big)
  \bigg]\bigg\}\,,\nonumber
\end{align}
where $\lambda_\rho = 2\alpha_s\beta_0\log(\tfrac{1}{\rho})$,
$\lambda_c = 2\alpha_s\beta_0\log(\tfrac{1}{\zcut})$ and
$\lambda_B=-2\alpha_s\beta_0B_i$ (associated with hard-collinear
splittings). and $W(x)=x\log(x)$,
$V(x)=\tfrac{1}{2}\log^2(x)+\log(x)$.
The expression above is computed using a two-loop running coupling in
the CMW scheme~\cite{Catani:1990rr}, and $\alpha_s$ is taken at the
hard scale $p_tR$.
The results for $R'$ can be straightforwardly obtained by taking a
derivative of the above expression wrt $\log(\tfrac{1}{\rho})$ and the
plain jet radiators are obtained by taking either $\beta$ to $\infty$
or $\zcut$ to $0$.

For $R_\text{SD}(\rho\tilde\tau;z_a)$, Eq.~(\ref{eq:Rrhotautildesd}),
we need two further ingredients: the possible extra contribution from
$\theta>\theta_a$ (and $z<\zcut\theta^\beta$, since the rest is
already included in the expression above), and the contribution from
secondary emissions.
Introducing
\begin{align}
& \delta R_\beta(\lambda_\text{top},\lambda_\text{bot}) = \frac{C_i}{2\pi\alpha_s\beta_0^2}\bigg\{
     \bigg[
    \frac{W(1-\lambda_\text{top})}{1+\beta}+W(1-\lambda_\text{bot})
    -\frac{2+\beta}{1+\beta}W\Big(1-\frac{\lambda_\text{top}+(1+\beta)\lambda_\text{bot}}{2+\beta}\Big)
     \bigg]\nonumber\\
&
\quad  +\frac{\alpha_s \beta_1}{\beta_0} \bigg[
    \frac{V(1-\lambda_\text{top})}{1+\beta}+V(1-\lambda_\text{bot})
    -\frac{2+\beta}{1+\beta}W\Big(1-\frac{\lambda_\text{top}+(1+\beta)\lambda_\text{bot}}{2+\beta}\Big)
  \bigg]\nonumber\\
&
\quad -\frac{\alpha_s K}{2\pi}\bigg[
    \frac{\log(1-\lambda_\text{top})}{1+\beta}+\log(1-\lambda_\text{bot})
    -\frac{2+\beta}{1+\beta}\log\Big(1-\frac{\lambda_\text{top}+(1+\beta)\lambda_\text{bot}}{2+\beta}\Big)
  \bigg]\bigg\}\,\Theta(\lambda_\text{bot}>\lambda_\text{top})\,,\nonumber
\end{align}x
we can write the ``extra triangle'' and secondary contributions as
\begin{align}
  R_{\text{SD,extra}}(\rho\tau,z_a)
  & = \delta R_\beta\Big(\lambda_c+(\beta+1)\frac{\lambda_\rho-\lambda_a}{2},
                        \frac{\lambda_\rho-\lambda_a}{2}+\lambda_\tau\Big), \\
  R_{\text{secondary}}(\rho\tau,z_a)
  & = \delta R_\beta\Big(\frac{\lambda_\rho+\lambda_a}{2},
                       \frac{\lambda_\rho+\lambda_a}{2}+\lambda_\tau\Big),
\end{align}
with $\lambda_\tau=2\alpha_s\beta_0\log(1/\tau)$ and $\lambda_a=2\alpha_s\beta_0\log(1/z_a)$.

\section{The multiple-emission function $f_{\text{ME}}$}\label{app:fme}

In practice, $f_{\text{ME}}(x,R')$ can be computed
analytically for $x\le 1$, and $1<x\le 2$ and we have managed to
reduce it to a single integration at least for $2<x\le 4$:
\begin{align}
f_{\text{ME}}(x,R')
 & \oset[0.12cm]{x\le1}{\quad=\quad} x^{R'-1}, \\
 & \oset[0.12cm]{1<x\le 2}{\quad=\quad} x^{R'-1}
   \bigg[1-\Big(\frac{x-1}{x}\Big)^{R'}
    {}_2F_1\Big(R',1,1+R',\frac{x-1}{x}\Big)\bigg],\nonumber\\
 & \oset[0.12cm]{2<x\le 3}{\quad=\quad} f_{\text{ME}}(2,R')
   + R'^2 \int_0^{x-2}du\,\frac{u^{R'-1}}{x-u}\log(x-1-u),\nonumber\\
 & \oset[0.12cm]{3<x\le 4}{\quad=\quad} f_{\text{ME}}(3,R')
   + R'^3 \int_0^{x-3}du\,\frac{u^{R'-1}}{x-u}\Big[\text{Li}_2\Big(\frac{1}{x-1-u}\Big)+\frac{1}{2}\log^2(x-1-u)-\frac{\pi^2}{12}\Big]\nonumber
\end{align}

In general, we write $f_{\text{ME}}(x,R')$ as an inverse Mellin
transform, which is what we have used for $x>4$:
\begin{equation}
f_{\text{ME}}(x,R') = \Gamma(R') \oint \frac{d\nu}{2i\pi}
e^{\nu x}
\exp\bigg\{\frac{R'}{2}\text{Ei}(-\nu)\bigg[\log(-\nu)-\log\bigg(\!\!-\frac{1}{\nu}\bigg)\bigg]\bigg\}.
\end{equation}

\section{Subleading contributions from similar angles}\label{app:subleading}

In this Appendix, we investigate the difference between the
${\cal{O}}(\alpha_s^2)$ fixed-order predictions and our analytic
expressions for the $\tau_{21}$ distribution in the shoulder region,
$\tau_{21}\lesssim \tfrac{1}{2}$, and trace it back to a subleading
contribution in the region where two emissions have similar angles.
To show this, we work at small jet radius and use the framework of the
integration over the triple-collinear splitting function.
At ${\cal{O}}(\alpha_s^2)$, a jet is made of 3 partons of momentum
fractions $z_i$ and pairwise angles $\theta_{ij}$ with $i,j=1,2,3$,
constrained so that
$z_1z_2\theta_{12}^2+z_1z_3\theta_{13}^2+z_2z_3\theta_{23}^2=\rho$.
For simplicity, we focus on the $C_F^2$ term, as the
other contributions are subleading in $\log(\rho)$. We can then assume
that particles 1 and 2 are gluons and particle 3 is a quark.

The expression for $\tau_2$ for the minimal axes can be obtained by
minimising over all possible partitions of the jet and can be written
as
\begin{align}\label{eq:tau21-3p-minimal}
  \tau_{2}^{\text{(min)}} = \text{min}\Big(
  \frac{z_1z_2}{z_1+z_2}\theta_{12}^2,
  \frac{z_1z_3}{z_1+z_3}\theta_{13}^2,
  \frac{z_2z_3}{z_2+z_3}\theta_{23}^2\Big)\qquad\qquad\text{[minimal]}.
\end{align}

\begin{figure}[!t]
  \includegraphics[width=0.48\textwidth,page=1]{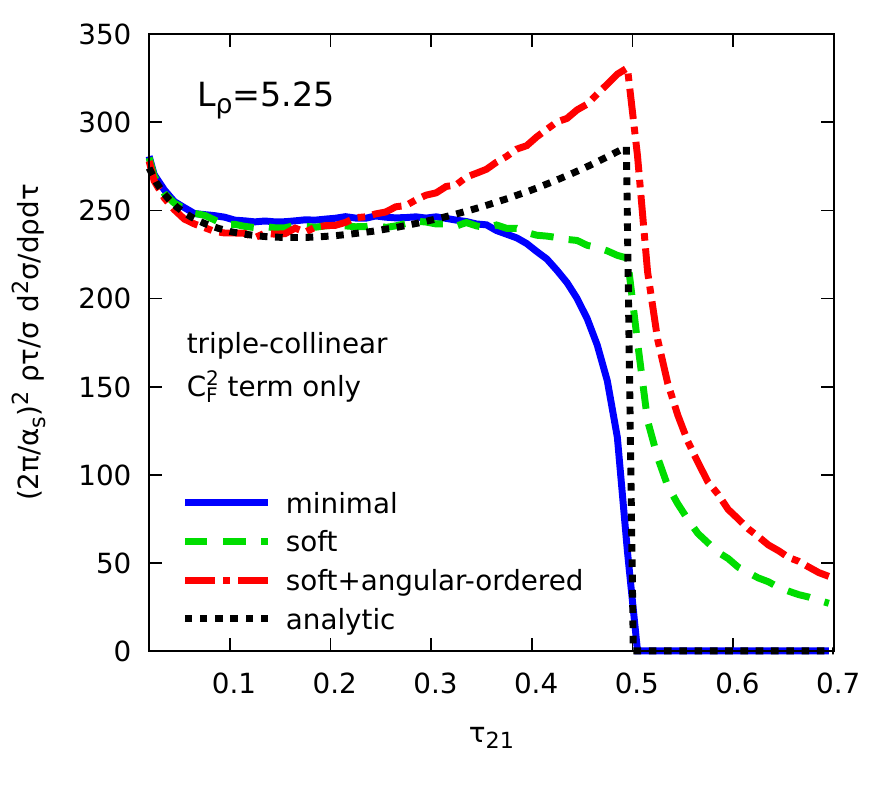}%
  \hfill%
  \includegraphics[width=0.48\textwidth,page=2]{figs/test-convergence.pdf}%
  \caption{$\tau_{21}$ distributions at ${\cal{O}}(\alpha_s^2)$ in the
    triple-collinear limit, obtained with different approximations for
    $\tau_2$. See text for details.} \label{fig:convergence}
\end{figure}

Our leading-logarithmic expression, Eq.~(\ref{eq:tau21}), is obtained
from $\tau_{2}^{\text{(min)}}$ by applying two approximations.
Firstly, logarithms of $\rho$ come from soft emissions,
$z_{1,2}\ll 1$, $z_3\approx 1$, yielding
\begin{align}\label{eq:tau21-3p-soft}
  \tau_{2}^{\text{(soft)}} = \text{min}\Big(
  \frac{z_1z_2}{z_1+z_2}\theta_{12}^2,
  z_1\theta_{13}^2,
  z_2\theta_{23}^2\Big)\qquad\qquad\text{[soft]},
\end{align}
with $\theta_{13}\approx \theta_1$ and $\theta_{23}\approx \theta_2$.
Secondly, if each emission comes with a logarithm of $\rho$, they can be
taken as strongly ordered in angles meaning
$\frac{z_1z_2}{z_1+z_2}\theta_{12}^2\approx
\text{max}(z_1\theta_1^2,z_2\theta_2^2)$ and therefore
\begin{align}\label{eq:tau21-3p-soft-ang-ord}
  \tau_{2}^{\text{(soft+ang.-ordered)}} = \text{min}\Big(
  z_1\theta_{13}^2,
  z_2\theta_{23}^2\Big)\qquad\qquad\text{[soft+ang.-ordered]},
\end{align}
which is to all practical purposes the expression (\ref{eq:tau21}) we use
throughout this paper.

In Fig.~\ref{fig:convergence}, we plot results obtained by integrating
the triple-collinear splitting function for the plain jet, with
$\tau_{21}$ computed using the three definitions above, and compare
the results with our analytical formula.
The striking feature here is that the above approximations mostly
affect the region close to $\tau_{21}=\tfrac{1}{2}$, meaning that
subleading logarithmic corrections are expected to have a
non-negligible impact in this region for reasonable values of
$\log(1/\rho)$.

It is helpful to discuss in a bit more details the differences
associated with the soft and angular-ordered approximations.
For the soft approximation, we see in Fig.~\ref{fig:lund}
(right) that the correction indeed only affect a region of finite
width at large $z_2$. This therefore gives at most a constant upon
integration over $z_2$, subleading compared to the $\log(1/\rho)$ one
would obtain from the integration in the soft limit.
Interestingly, the difference between the minimal axes and the soft
approximation appears mostly in the region above
$\tau_{21}=\tfrac{1}{2}$, where we also see differences between the
minimal and generalised-$k_t$ choices of axes. Although we have not
explicitly checked that, the value of $\tau_{21}$ generalised-$k_t$ is
likely affected by factors of $1-z$ in that region, due to
differences between a pairwise mass $z_iz_j\theta_{ij}^2$ and the
generalised-$k_t$ distance $\text{min}(z_i,z_j)\theta_{ij}^2$.

Next, we want to show explicitly that the contribution coming from
emissions of similar angles, \ie using $\tau_{2}^{\text{(soft)}}$
instead of $\tau_{2}^{\text{(soft+ang.-ordered)}}$, also leads to a
subleading correction.
This is particularly interesting because, from
Fig.~\ref{fig:convergence}, it appears to be the main contribution
driving the shoulder effect for $\tau_{21}\lesssim \tfrac{1}{2}$.
For simplicity, let us consider the case of the cumulative
distribution $H(\rho,>\tau)$ with two emissions ``1'' and ``2'', and
look at the contribution coming from the integration over emission
``2'', with $\rho_2<\rho_1$ for a fixed $\rho$ and $\theta_1$. This
can be written as\footnote{The $\rho-\rho_2$ denominator comes from
  the integration over $\rho_1$ with the constraint $\rho_1+\rho_2=\rho$.}
\begin{equation}
  I_2 = \int_0^{\rho/2} \frac{\rho\, d\rho_2}{\rho_2(\rho-\rho_2)}
  \int_{\rho_2}^1\frac{d\theta_2^2}{\theta_2^2}\int_0^{2\pi}\frac{d\phi}{2\pi}
  \Theta\Big(\text{min}\Big(\frac{z_1z_2}{z_1+z_2}\theta_{12}^2,\rho_2\Big)>\rho\tau\Big),
\end{equation}
where we use the soft approximation,
Eq.~(\ref{eq:tau21-3p-soft}), and $\theta_{12}^2=\theta_1^2-2\theta_1\theta_2\cos\phi+\theta_2^2$.
We can write $I_2$ as a ``leading'' contribution coming from the
approximation in~(\ref{eq:tau21-3p-soft-ang-ord}) and a correction:
\begin{align}
  I_2 & = I_{2,\text{leading}} -\delta I_2,\label{eq:I2-split}\\
  I_{2,\text{leading}}& = \int_0^{\rho/2} \frac{\rho\, d\rho_2}{\rho_2(\rho-\rho_2)}
  \int_{\rho_2}^1\frac{d\theta_2^2}{\theta_2^2}\int_0^{2\pi}\frac{d\phi}{2\pi}
      \Theta(\rho_2>\rho\tau)
  = \log\Big(\frac{1-\tau}{\tau}\Big)\log\Big(\frac{1}{\rho}\Big)+\text{const.},\label{eq:I2-leading}\\
  \delta I_2 & =\int_0^{\rho/2} \frac{\rho\, d\rho_2}{\rho_2(\rho-\rho_2)}
  \int_{\rho_2}^1\frac{d\theta_2^2}{\theta_2^2}\int_0^{2\pi}\frac{d\phi}{2\pi}
  \Theta\Big(\text{min}\Big(\frac{z_1z_2}{z_1+z_2}\theta_{12}^2,\rho_2\Big)<\rho\tau<\rho_2\Big).\label{eq:delta-I2}
\end{align}
Here, $I_{2,\text{leading}}$ is the leading contribution we
compute to all orders in this paper and $\delta I_2$ is a
correction. We want to show that $\delta I_2$ is subleading, \ie that
it does not come with any $\log(\rho)$ enhancement.
For that it is sufficient to show that the integration over
$\theta_2$, which is at the origin of the $\log(1/\rho)$ in
$I_{2,\text{leading}}$, now contributes at most to a constant
in $\rho$.
For the constraint in~(\ref{eq:delta-I2}) to be non-zero, we need
$\tfrac{z_1z_2}{z_1+z_2}\theta_{12}^2<\rho_2$ and $\cos(\phi)<1$ from
which we get
$\tfrac{\theta_2}{\theta_1}>\tfrac{\rho-2\rho_2}{2(\rho-\rho_2)}$.
Since the right hand side is a number, the limit of small $\theta_2$
does not give a large logarithm of $\rho$.\footnote{Note however that
  it would be interesting to further investigate this contribution as
  $\tau$ approaches $\tfrac{1}{2}$ where $\rho-2\rho_2$ can approach
  0. In this case the integration over $\phi$ would still be
  suppressed by a power of $\tfrac{\theta_2}{\theta_1}$ but it might
  be sufficient to discuss the transition around $\tau=\tfrac{1}{2}$.}

\begin{figure}[!t]
  \centerline{\includegraphics[width=0.48\textwidth]{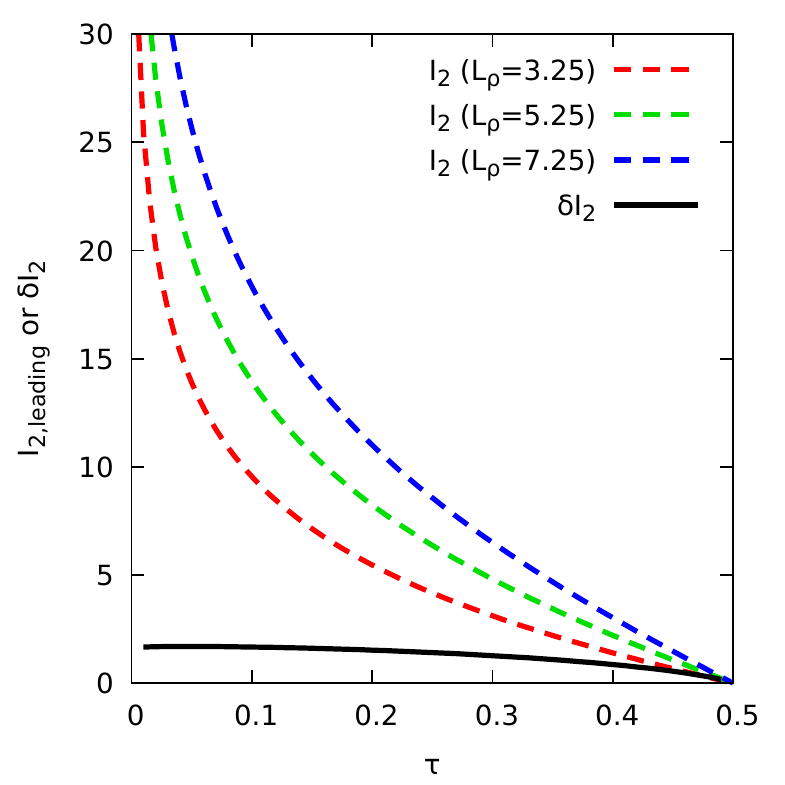}}%
  \caption{Contribution of the subleading $\delta I_2$ contribution,
    compared to the leading $I_{2,\text{leading}}$ contribution for
    different values of $L_\rho=\log(1/\rho)$.} \label{fig:subleading}
\end{figure}

In the limit of large $\theta_2$, we can rewrite the constraint as
$\tau<\tfrac{\rho_2}{\rho}<\tau(1+2\tfrac{\theta_1}{\theta_2}\cos\phi+{\cal{O}}(\tfrac{\theta_1^2}{\theta_2^2}))$.
The integration over $\rho_2$ therefore brings an extra factor
$\tfrac{\theta_1}{\theta_2}$ suppressing the large-$\theta_2$
contribution. This corresponds to the decrease towards a ratio of 1 at
large $\theta_2$ in Fig.~\ref{fig:lund}. Altogether, this implies that
$\delta I_2$ does not have any $\log(1/\rho)$ enhancement.  To further
illustrate this point, we plot $\delta I_2$ in
Fig.~\ref{fig:subleading}, compared to the leading contribution
$I_{2,\text{leading}}$. We obtain this by numerically integrating
Eqs.~(\ref{eq:I2-leading}), setting the limits of the $\theta_2$
integration to $\pm\infty$ so that it becomes independent of $\rho$,
and keeping only the leading $\log(\rho)$ contribution in
$I_{2,\text{leading}}$. We clearly see on this plot that the
$\delta I_2$ contribution has a relatively larger impact as $\tau$
increases.

\clearpage

\bibliographystyle{UTPstyle}
\bibliography{bib}
\end{document}